\documentclass[a4paper,12pt]{article}

\usepackage[normalem]{ulem}
\useunder{\uline}{\ul}{}

\usepackage{paralist}
\usepackage{comment}
\usepackage[utf8]{inputenc} 
\usepackage[T1]{fontenc}
\usepackage[stable]{footmisc}
\usepackage[margin=1in]{geometry}
\usepackage{graphicx} 
\usepackage{cancel}

\usepackage[toc,page]{appendix}
\usepackage{booktabs} 
\usepackage{array}    
\usepackage{paralist} 
\usepackage{verbatim} 
\usepackage[table]{xcolor}
\usepackage{multirow}
\usepackage{rotating}
\usepackage{lscape}
\usepackage{tabularx}
\usepackage{booktabs}
\usepackage{amsthm}
\usepackage{setspace}
\usepackage{amssymb,mathtools,amsthm,mathrsfs}
\usepackage{tikz}
\usepackage{ctable}
\usetikzlibrary{fit}
\usepackage{graphicx}
\usepackage{caption}
\usepackage{subcaption}
\usepackage{bm}
\usepackage{float}
\usepackage{bbm}
\usepackage{natbib}
\usepackage[flushleft]{threeparttable}
\usepackage{makecell}

\usepackage{setspace}
\AtBeginEnvironment{table}{\singlespacing}

\newcommand{\R}{\mathbb{R}}

\newcommand{\E}{\mathbb{E}}
\newcommand{\bs}{\boldsymbol}

\newcommand\Tstrut{\rule{0pt}{2.9ex}}         
\newcommand\Bstrut{\rule[-1.2ex]{0pt}{0pt}}   

\title{Forecasting Large Realized Covariance Matrices: The Benefits of Factor Models and Shrinkage \thanks{We thank Michael Wolf for insightful comments and for providing us with the Matlab code to estimate the NL-DCC, NL-BEKK and AFM1-DCC-NL models. We owe Gialuca DeNard for providing codes for the IDR-DCC-NL model. We are also thankful to the seminar participants at Aarhus, Barcelona GSE, Erasmus Research Institute of Management, ETH Zurich, IMPA, University of Padua, Tinbergen Institute, the Third International Workshop in Financial Econometrics (October 2017), Quantitative Finance and Financial Econometrics Conference (May 2018), IAAE International Association for Applied Econometrics Conference (June 2018), SoFIE (June 2018) and SBFin (July 2018) for providing us with many comments and suggestions. We are also in debt with Tim Bollerslev, Dick van Dijk, Marcelo Fernandes, René Garcia, Diogo Guillen, Roxana Halbleib, Hedibert Lopes, Bradley Paye and Rogier Quaedvlieg for comments and suggestions. Alves: rafael.alves1193@gmail.com. Brito: diego.debrito.vaa5fz@statefarm.com. Medeiros: marcelom@illinois.edu. Ribeiro: ruymr@insper.edu.br.}}

\author{ {Rafael P. Alves} \\ \small{Duke University} \and {Diego S. de Brito} \\ \small{State Farm} \and {Marcelo C. Medeiros} \\ \small{The University of Illinois at Urbana-Champaign} \and {Ruy M. Ribeiro} \\ \small{Insper}}

\bigskip \bigskip \bigskip

\date{\small{First version: October 2017 \\ This version: February 2023}}

\begin{document}

\maketitle

\doublespacing

\begin{abstract}
We propose a model to forecast large realized covariance matrices of returns, applying it to the constituents of the S\&P 500 daily. To address the curse of dimensionality, we decompose the return covariance matrix using standard firm-level factors (e.g., size, value, and profitability) and use sectoral restrictions in the residual covariance matrix. This restricted model is then estimated using vector heterogeneous autoregressive (VHAR) models with the least absolute shrinkage and selection operator (LASSO). Our methodology improves forecasting precision relative to standard benchmarks and leads to better estimates of minimum variance portfolios.

\end{abstract}

\pagebreak

\section{Introduction}

This paper aims to construct models based on economically motivated factor decompositions and shrinkage methods to forecast large-dimensional, and time-varying realized measures of daily covariance matrices of returns on financial assets. Realized measures of a covariance matrix are estimates, based on intraday returns, of the integrated covariance matrix of a multivariate diffusion process. One example of such an estimator used in this paper is the composite realized kernel method recently introduced by \citet{lunde2016econometric}. Our proposed model is evaluated in terms of its forecasting ability and, more importantly, several performance measures in a conditional mean-variance portfolio allocation problem.

Modeling and forecasting the covariance matrix of financial assets are essential for portfolio allocation and risk management. Moreover, it is an established empirical fact that (conditional) covariance matrices vary considerably over time. A natural way to model such dynamics is to either use multivariate generalizations of the ARCH/GARCH family of models, as proposed by \citet{BEEK1995} or \citet{EngleDCC2002}, or model directly a given realized measure of the covariance matrices by usual multivariate time-series models, as in \citet{Bauer2011}, \citet{Chiriac2011} or \citet{golosnoy2012conditional}. This is motivated by the close connection between the conditional covariance matrix and the integrated covariance of a multivariate diffusion process.

However, when the number of assets increases, the amount of parameters to be estimated becomes very large. For instance, for a covariance matrix of $N$ assets, there are $N(N+1)/2$ distinct entries to be modeled. If a vector autoregressive specification of order $p$, VAR($p$), is used, then the total number of parameters will be $(N+1)(p+1)/2$. Therefore, the curse of dimensionality precludes the application of the above-referenced methods to moderately large covariance matrices, and most of the previous studies in the literature focused on sets of less than ten assets.

More recently, based on the advances of modern statistical tools to handle high-dimensional models, new alternatives have been proposed in the literature to address a large number of assets. \citet{callot2017modeling} advocate the use of the least absolute shrinkage and selection operator (LASSO) of \citet{tibshirani1996regression} and the adaptive LASSO of \citet{zou2006adaptive} to model the dynamics of the realized covariance of the constituents of the Dow Jones index by a large dimensional VAR model. However, their modeling strategy cannot handle sets of assets much larger than the 30 used in their paper. \citet{engle2017large} combine nonlinear shrinkage with the DCC-GARCH model and put forward a methodology in which the dynamics of large latent conditional covariance matrices can be modeled. Their approach makes the estimation of multivariate GARCH-type models feasible even in large-dimensional cases. Nevertheless, the authors did not consider realized measures in their models.

In this paper, we propose a method that can be applied to a large set of assets with the difference that we consider realized covariance measures rather than latent ones, as in \citet{engle2017large}. However, it is important to highlight that our method relies on realized measures of the integrated covariance matrix. Although our methodology can be applied to thousands of assets, illiquid assets may cause severe problems for the integrated covariance matrix estimation. For instance, although the estimator by \citet{lunde2016econometric}, which is used in this paper, is scalable as it is composed of 2x2 matrix estimates, an illiquid stock will cause a poor estimate of that stock’s covariance with other stocks. Nevertheless, the estimates of covariances between other liquid stocks will not be affected.

We provide empirical evidence that our proposed model drastically improves the forecasting ability and produces portfolios with better performance measures. We apply our model to some of the constituents of the S\&P 500 using daily information on the realized covariance matrix. The models based on our framework can improve the forecasts' quality compared to the random walk benchmark by reducing the $\ell_2$-forecast error in 15\%. On the other hand, models based on latent volatility are the worst-performing specification regarding forecasting ability. Furthermore, our models deliver a significant reduction in the volatility of minimum variance portfolios, as the annualized standard deviation of returns declines by more than 25\% relative to the second-best alternative in the literature and to nearly 40\% versus the random walk alternative. We also observe a decline in realized portfolio risk when considering longer investment horizons of up to one month and when we introduce portfolio constraints.

Our model applies a methodology similar to \citet{callot2017modeling} to the elements of the factor decomposition of realized covariance matrices. Hence, the daily covariance matrix of returns is first written into a covariance matrix for a low-dimensional set of factors plus an idiosyncratic covariance matrix that is (almost) block diagonal. We consider economically motivated factors widely used in the finance literature based on firm-level characteristics such as size, value, and profitability. The dynamics of the variances and covariances of the factors are modeled by a VHAR model estimated with LASSO. To guarantee positive definiteness of the forecasts, we apply the log matrix transformation of \citet{chiu1996matrix}. The daily factor loadings (``betas'') are computed with high-frequency data. We show that these loadings are time-varying and exhibit a high degree of long-range dependence, and we model their dynamics by a HAR specification. Finally, the dynamics of each block of the idiosyncratic covariance matrix is modeled by a restricted autoregressive model, also estimated by LASSO.

In this paper, we do not address the interesting challenges involved in the construction of realized measures (e.g., how to address microstructure noise), especially in the multivariate case (e.g., asynchronicity in transactions of different assets, which bias covariance estimations toward zero). We focus solely on modeling and forecasting a large realized covariance matrix of returns on hundreds of financial assets estimated elsewhere.

The remainder of the paper is organized as follows. Sections \ref{S:Model} and  \ref{section_forecasting_methodology} describe the proposed model and forecasting framework. In Section \ref{S:Data}, we present the data and show a descriptive analysis. Forecasting results and portfolio analysis are presented, respectively, in Sections \ref{S:Results} and \ref{sec_portfolio}. Finally, Section \ref{S:Conclusions} concludes the paper. Supplementary results are shown in the Appendix.

\section{Model}\label{S:Model}
We base our methodology on the fact that realized covariance matrices are highly persistent over time, which suggests the use of an autoregressive model of a large order $p$, usually larger than 20. Defining $\bs y_{t}=\textnormal{vech}(\bs\Sigma_{t})$, where vech is the half-vectorization operator returning a vector with the unique entries of $\bs\Sigma_{t}$, one possible specification is
\begin{equation}\label{autoregression_raw}
\bs y_{t}=\bs\omega + \sum_{i=1}^{p}\bs\Phi_{i}\bs y_{t-1}+\bs\epsilon_{t},
\end{equation}
where $\bs\epsilon_{t}$ is zero-mean random noise.

While this model is sensible for a small number of assets, the number of parameters grows quadratically as new assets are added (curse of dimensionality). To circumvent this limitation, \citet{callot2017modeling} propose using penalized regressions (LASSO) to address the large number of parameters. However, the direct use of penalized regressions is unfeasible for hundreds of assets.

To illustrate, assume that the representation in (\ref{autoregression_raw}) is used to model the covariance matrix for the constituents of the S\&P 500 index\footnote{In our application, we restrict the analysis to stocks that remain in the index for the entire period of our sample. This reduces the number of stocks to 430. Regarding the number of equations and potential predicting variables, our problem is still subject to the curse of dimensionality on the same order of magnitude as the illustration.} with 10 lags, that is, $p=10$. Since each matrix $\bs\Sigma_{t}$ has $N(N+1)/2$ unique entries, this configuration would result in $125,250$ equations with $10\times 125,250$ variables each, plus constants. In this case, estimation is unfeasible even with LASSO.

To reduce the dimensionality to a manageable one, we propose the use of a factor model as well as economic restrictions based on sector classifications and penalized regressions.

\subsection{Factor Model}

Following the factor model discussed in \citet{chamberlain1983arbitrage}, the excess return on any asset $i$, denoted $r_{i,t}$, satisfies
\begin{equation}
r_{i,t}^{e} = \beta_{i1,t}f_{1,t} + \cdots + \beta_{iK,t}f_{K,t} + \varepsilon_{i,t} = \bs\beta_{i,t}'\bs f_{t} +\varepsilon_{i,t},
\end{equation}
where $f_{1,t},\cdots, f_{K,t}$ are the excess returns of $K$ factors, $\beta_{ik,t}$, $k=1,\ldots, K$, are factor loadings for asset $i$, and $\varepsilon_{i,t}$ is the idiosyncratic error term. Note that the factor loadings are allowed to be variable over time. For $N$ assets, the set of equations can be written in matrix form:
\begin{equation} \label{eq_factor_model}
\bs r_{t}^{e} = \bs B_{t}'\bs f_{t} + \bs\varepsilon_{t},
\end{equation}
where $\bs B_{t}$ is a $K \times N$ matrix of loadings, $\bs r_{t}$ is an $N \times 1$ vector of excess returns, and $\bs\varepsilon_{t}$ is an $N \times 1$ vector of idiosyncratic errors. Throughout, we assume that $\E(\bs\varepsilon_{t}|\bs f_{t})=\bs 0$.
The factors used in this work are linear combinations of returns constructed solely with the assets considered here, i.e., long-short stock portfolios where stocks that are part of our sample are sorted on firm characteristics. In matrix form, for all $K$ factors,
\begin{equation}\label{factors_components}
\begin{gathered}
\begin{pmatrix}
f_{1,t}\\
\vdots\\
f_{K,t}
\end{pmatrix}=
\begin{pmatrix}
w_{1,1}  & \hdots &  w_{1,N} \\
\vdots   & \ddots & \vdots \\
w_{K, 1} & \hdots & w_{K, N}
\end{pmatrix}
\begin{pmatrix}
r_{1,t}\\
\vdots\\
r_{N,t}
\end{pmatrix} \\ \textnormal{or} \\
\bs f_{t}=\bs W'\bs r_{t},
\end{gathered}
\end{equation}
where weights are calculated based on accounting and market information, as we describe in Section \ref{S:Data}.

\subsection{Covariance Decomposition}
This section describes how we decompose the realized covariance matrix of returns for all assets into a factor covariance matrix and a residual covariance matrix. Let $\bs\Sigma_{t}$ denote the realized covariance matrix of returns at time $t$, that is, $\bs \Sigma_{t} =\textnormal{cov}(\bs r_{t})$. By using equation (\ref{eq_factor_model}) and the assumption $\E(\bs \varepsilon_{t}|\bs f_{t})=\bs 0$, we have

\begin{equation}\label{covariance_decomposition}
\bs\Sigma_{t}=\textnormal{cov}(\bs B_{t}'\bs f_{t})+\textnormal{cov}(\bs\varepsilon_{t})=\bs B_{t}'\bs\Sigma_{f,t}\bs B_{t}+\bs\Sigma_{\bs\varepsilon,t},
\end{equation}
where $\bs B_{t}$ is a $K \times N$ matrix of loadings of $N$ assets on $K$ factors, $\bs\Sigma_{f,t}$ is the $K \times K$ factor covariance matrix and $\bs\Sigma_{\bs\varepsilon, t}$ is the $N \times N$ residual covariance matrix, all at time $t$. Since each factor is a linear combination of returns, factor covariance matrices can be obtained by using equation (\ref{factors_components}) and the known values of $\bs\Sigma_{t}$, that is,

\begin{equation}
\bs\Sigma_{f,t}=\textnormal{cov}(\bs f_{t})=\textnormal{cov}(\bs W'\bs r_{t})=\bs W'\bs\Sigma_{t}\bs{W}.
\end{equation}

Factor loadings $\bs B_{t}$ are calculated using a similar procedure (see Appendix \ref{appendix_factor_loadings}), and the values of $\bs\Sigma_{\bs\varepsilon,t}$ are simply given by the difference $\bs\Sigma_{t}-\bs B_{t}'\bs \Sigma_{f,t}\bs B_{t}$. It is common to assume that $\bs\Sigma_{\bs\varepsilon,t}$ is diagonal, but we will be less restrictive in this work.

\section{Forecasting Methodology}\label{section_forecasting_methodology}
With the decomposition achieved in (\ref{covariance_decomposition}), we write the forecasting equation for the complete covariance matrix $\bs\Sigma_{t}$ by combining separate forecasts of $\bs B_{t}$, $\bs\Sigma_{f,t}$, and $\bs\Sigma_{\bs\varepsilon, t}$. That is,

\begin{equation}
\widehat{\bs\Sigma}_{t+1|t}=\widehat{\bs B}_{t+1|t}'\widehat{\bs\Sigma}_{f,t+1|t}\widehat{\bs B}_{t+1|t} + \widehat{\bs\Sigma}_{\bs\varepsilon, t+1|t}.
\end{equation}
\subsection{Forecasting $\bs\Sigma_{f,t}$}\label{forecasting_factors_covariance}

Since the number of factors is much smaller than the number of assets, one could propose using an unrestricted VAR model for the factor covariance matrix dynamics, as in equation (\ref{autoregression_raw}). Despite the reduction in dimensionality achieved by using a factor model, the number of parameters in this configuration is still quite large (note that each equation would have $K(K+1)/2+1$ parameters). To reduce this number to a more manageable one, we follow the heterogeneous autoregressive (HAR) model proposed by \citet{corsi2009simple}. In this model, the predictors are obtained from the simple average of daily data computed for different horizons (daily, weekly, and monthly). In our case, daily, weekly, and realized covariance matrices of factors are given by, respectively,
 \begin{equation}\label{factors_har}
\begin{gathered}
\bs\Sigma_{f,t}^{day}=\bs\Sigma_{f,t} \\
\bs\Sigma_{f,t}^{week}=\frac{1}{5}(\bs\Sigma_{f,t}+\bs\Sigma_{f,t-1}+\cdots+\bs\Sigma_{f,t-4})\\
\bs\Sigma_{f,t}^{month}=\frac{1}{22}(\bs\Sigma_{f,t}+\bs\Sigma_{f,t-1}+\cdots+\bs\Sigma_{f,t-21}),
 \end{gathered}
 \end{equation}
where $\bs y_{f,t}=\textnormal{vech}(\bs\Sigma_{f,t})$, $\bs y_{f,t}^{day}=\textnormal{vech}(\bs\Sigma_{f,t}^{day})$, $\bs y_{f,t}^{week}=\textnormal{vech}(\bs\Sigma_{f,t}^{week})$, and
$\bs y_{f,t}^{month}=\textnormal{vech}(\bs\Sigma_{f,t}^{month})$. Furthermore, the dynamic process for $y_{f,t}$ is defined as
\begin{equation}\label{factors_dynamics}
\bs y_{f,t}=\bs\omega + \bs\Phi_{day}\bs y_{f,t-1}^{day} + \bs\Phi_{week}\bs y_{f,t-1}^{week} + \bs\Phi_{month}\bs y_{f,t-1}^{month} + \bs\epsilon_{t},
\end{equation}
where $\bs\Phi_{day}$, $\bs\Phi_{week}$, and $\bs\Phi_{month}$ are $M \times M$ matrices, where $M=K(K+1)/2$ is the number of unique entries on the factor covariance matrix. $\bs\omega$ is an $M\times 1$ vector of constants.

\subsubsection{LASSO and adaLASSO}

Due to the high number of parameters in equation (\ref{factors_dynamics}), direct estimation with ordinary least squares (OLS) could result in overfitting, harming the precision of model forecasts. LASSO shrinks these estimates by imposing a penalty related to the magnitude of the coefficients, shrinking them towards zero (\citet{tibshirani1996regression}).  This methodology has been shown to almost always provide a higher out-of-sample forecasting precision, while the reduced number of predictors makes interpreting the model easier.

We estimate (\ref{factors_dynamics}) equation by equation. Consider a sample size of $T$, and let $\bs{Z}_{t} = \left(1, \bs{y}_{f,t-1}^{day'},\bs{y}_{f,t-1}^{week'},\bs{y}_{f,t-1}^{month'}\right)'\in\R^{3m + 1}$ be the vector of explanatory variables and $\bs Z=(\bs Z_{T},\ldots,\bs Z_1)'$ be the $T \times (3m+1)$ matrix of covariates. Let $\bs y_{f,i} = (y_{f,T,i},\ldots, y_{f,1,i})'\in\R^T$ be the vector of observations on the $i$th equation of (\ref{factors_dynamics}), $i=1,\ldots,M$, and $\bs\epsilon_{f,i}=(\epsilon_{f,T,i},\ldots,\epsilon_{f,1,i})'$ be the corresponding vector of error terms. With $\bs \gamma_{f,i}=(\omega_{i},\bs\beta_{i}')'$ being the $3M+1$ vector of true parameters for equation $i$, one can write
\begin{equation}
\bs y_{f,i}=\bs Z\bs\gamma_{f,i}+\bs\epsilon_{f,i},\quad\quad i=1,\ldots,M,
\end{equation}
where each vector $\bs\gamma_{f,i}$ is then estimated by minimizing
\begin{equation}\label{eq_lasso}
L(\bs\gamma_{f,i})=\frac{1}{T}||\bs y_{f,i}-\bs Z\bs\gamma_{f,i}||^2+2\lambda_{T}||\bs\beta_{i}||_{\ell_1}, \quad \quad i=1,\ldots,M.
\end{equation}
The penalty parameter $\lambda_{T}$ determines how much penalization is imposed on the size of the coefficients. In our setup, the value of $\lambda_{T}$ is determined by minimizing the Bayesian information criterion (BIC). For equation $i$ and penalty parameter $\lambda$, the BIC is given by
\begin{equation}
\textnormal{BIC}_{i}(\lambda)=T\times \log(\widehat{\bs\epsilon}_{\lambda,i}'\widehat{\bs\epsilon}_{\lambda,i}) + \sum_{j=1}^{3M}\mathbbm{1}(\widehat{\beta}_{ij}^{\lambda} \neq 0)\log(T), \quad \quad i=1,\ldots,M,
\end{equation}
where $\widehat{\bs\epsilon}_{\lambda,i}$ is the estimated vector of error terms for penalty $\lambda$.

After obtaining $\widehat{\bs\gamma}_{f,i}$ from equation (\ref{eq_lasso}), the one-step-ahead forecast is given by
\begin{equation}\label{eq_forecasts}
\widehat{y}_{f,T+1,i}=\widehat{\bs\gamma}_{f,i}' \bs{Z}_{T}, \quad \quad i=1,\ldots,M,
\end{equation}
which can be used to provide $\widehat{\bs \Sigma}_{f,T+1}$.

We also consider the adaptive version of  LASSO (adaLASSO), as in \citet{zou2006adaptive}. adaLASSO is a two-step procedure that estimates the model via LASSO in the first step and excludes the variables classified as zero from the second step. The second step estimates the parameters with a slightly modified objective function that considers the size of the parameters estimated in the first step. To illustrate, consider the set of indices of the coefficients in the $i$th equation that are different from zero in the first step: $J(\widehat{\bs\beta}_{i})=\{j \in R^{3M}: \widehat{\beta}_{i,j} \neq 0\}$. adaLASSO estimates the vector of parameters $\bs\gamma_{f,i}$ by minimizing the following objective function:

\begin{equation}
L(\bs\gamma_{f,i})=\frac{1}{T}||\bs y_{f,i}-\bs Z\bs\gamma_{f,i}||^2 + 2\lambda_{T} \sum\limits_{j \in J(\hat{\bs\beta_{i}})} \frac{|\beta_{i,j}|}{|\hat{\beta}_{i,j}|}, \quad \quad i=1,\ldots,M.
\end{equation}

As before, the penalty parameter $\lambda_T$ is chosen by minimizing the BIC, and the one-step-ahead forecasts are computed as in equation (\ref{eq_forecasts}).

\subsection{Forecasting $\bs B_{t}$}

Instead of using unconditional loadings on factors, we assume that betas change daily and have long-range dependence. In Appendix \ref{appendix_betas_long_memory}, we show the distribution of the estimates of the fractional integration parameter for the beta series. The results show that these series present high persistence, similar to what is observed in realized covariance data. Based on this evidence and to maintain consistency with the rest of our methodology, we forecast each series of betas using HAR models. For each element of $\bs B_{t}$, we have
\begin{equation}\label{beta_dynamics}
\beta_{k,i,t}=\omega + \Phi_{day}\bs \beta_{k,i,t-1}^{day} + \Phi_{week}\bs \beta_{k,i,t-1}^{week} + \Phi_{month}\bs \beta_{k,i,t-1}^{month} + \epsilon_{k,i,t},
\end{equation}
where $\beta_{k,i,t}$, the entry $k\times i$ in the matrix $\bs B_{t}$,  is the loading of stock i on factor k at date t. We estimate the coefficients $\Phi_{day}$, $\Phi_{day}$, and $\Phi_{day}$ by OLS.

\subsection{Forecasting $\bs\Sigma_{\bs\varepsilon,t}$}\label{subsection_sigma_epsilon}
Since the residual covariance matrix dimension is $N \times N$, the curse of dimensionality remains a concern when forecasting $\bs\Sigma_{\bs\varepsilon,t}$. A factor model may not explain all the covariances between assets depending on the number and choice of factors, implying the covariance matrix of the residuals is not diagonal. Instead of imposing the common restriction that $\bs\Sigma_{\bs\varepsilon_,t}$ is diagonal (as in \citet{fan2008high}), we consider a less restrictive assumption that stocks in the same sector may still co-vary even after controlling for the standard factors used in the finance literature.

We assume that $\bs\Sigma_{\bs\varepsilon,t}$ is block diagonal, where blocks are defined by industry classification. After controlling for factor exposure, we assume there is no contemporaneous correlation between assets in different sectors. We argue that this assumption is not too strong in Section \ref{subsection_residual_covariance}, where we analyze residual correlations between stocks in different sectors.

We assume that each asset belongs to an industrial sector $s$, where $S$ is the total number of sectors and $S << N$. Without loss of generality, we can order assets so that the residual covariance matrix has sector covariance matrices near its diagonal:

\begin{equation}
\bs\Sigma_{\bs\varepsilon,t} =
\begin{pmatrix}
\bs\Sigma_{\bs\varepsilon,t}^{1} &  &\\
& \ddots  &  \\
&  & \bs\Sigma_{\bs\epsilon,t}^{S}
\end{pmatrix}.
\end{equation}

Depending on the selected industry classification, the number of assets in each group $s$ (called $N^{s}$) can still be quite large, as each block has $M^{s}=N^{s}(N^{s}+1)/2 $ unique elements.

Our second assumption is that the dynamics of each block $\bs\Sigma_{\bs\varepsilon,t}^{s}$ depend only on the elements of the same block at $t-1$, i.e., only $\bs\Sigma_{\bs\varepsilon,t-1}^{s}$ predict values in $\bs\Sigma_{\bs\varepsilon,t}^{s}$. To further simplify this model, we also consider an additional restriction that only past variances matter in this prediction. This last assumption relies on the previous evidence in \citet{callot2017modeling} that LASSO more frequently selects past variances as good predictors of covariance and variance terms than past covariances.

With the notation $\bs y_{\epsilon,t}^{s}=\textnormal{vech}(\bs\Sigma_{\bs\varepsilon,t}^{s})$, we have

\begin{equation}\label{error_block_equations}
\bs y_{\bs\varepsilon,t}^{s}=\bs\omega_{\epsilon}^{s}+\bs\Phi^{s}\bs\Lambda_{\bs\varepsilon,t-1}^{s}+\bs u_{\bs\varepsilon,t}^{s}, \quad \quad s=1,\ldots,S,
\end{equation}
where $\bs\omega_{\bs\varepsilon}^{s}$ is an $M^{s} \times 1$ vector of intercepts, $\bs\Phi^{s}$ is an $M^{s} \times N^{s}$ matrix of coefficients, $\bs\Lambda_{\bs\varepsilon,t-1}^{s} = \mathsf{diag}(\bs\Sigma_{\bs\varepsilon,t-1}^{s})$ is an $N^{s} \times 1$ vector of past variances, and $\bs u_{\bs\varepsilon,t}^{s}$ is the vector of errors.

The parameter estimation is done block by block. Each equation in (\ref{error_block_equations}) is estimated separately. The procedure is the same as used for the factor covariance matrix model: LASSO/adaLASSO regression equation by equation. We then regroup the one-day ahead forecast for each group, $\widehat{\bs\Sigma}_{\bs\varepsilon, T+1}^{s}$, to form the full residual covariance matrix forecast $\widehat{\bs\Sigma}_{\bs\varepsilon, T+1}$.

\section{Data and Descriptive Analysis}\label{S:Data}
\subsection{Realized Return Covariance Matrices}

The data consist of daily realized covariance matrices of returns for constituents of the S\&P 500 index. These matrices were constructed using 5-minute returns by the composite realized kernel method (discussed in \citet{lunde2016econometric} and provided by the authors). The full sample comprises all business days between January 2006 and December 2011. We consider companies that remained in the index and had balance sheet data available for the full sample period, resulting in a total of 430 stocks. With these considerations, the dataset consists of 1,495 daily $430 \times 430$ realized covariance matrices of returns.

\subsection{Data Cleaning}
Our sample spans the 2008 financial crisis and flash crashes from 2010 and 2011. To mitigate the effect of these extreme events, we perform a light cleaning on the realized covariance matrices, as in \citet{callot2017modeling}. For each day, we verify which (unique) entries are more than four standard deviations (of the series corresponding to that entry) away from their sample average up to then. If more than 25\% of the unique entries have extreme values according to this criterion, we flag that specific day. We replace the matrices corresponding to the flagged days by an average of the nearest ten preceding non-flagged matrices.

\subsection{Factors}
We construct each factor as the return time series for a long-short stock portfolio derived from individual sorts of underlying stocks on different signals. Since our approach uses the loading matrix $\bs W$ as an input to calculate factor covariance matrices and factor loadings, we could not use the widely available data on financial factors series of returns (such as on Kenneth French's website). Instead, we construct our versions of the standard factors, ranking our universe of stocks into the different signals to calculate the matrices $\bs W$ for our sample.

In total, we use seven factors widely used in the finance literature. Besides the market factor, we also consider size (SMB) and value (HML) (\cite{fama1993common}), gross profitability (\cite{novy2013other}), investment (\cite{lyandres2008new}), asset growth (\cite{cooper2008asset}), and accruals (\cite{sloan1996}). We report a detailed description of the construction of each factor in Appendix \ref{appendix_factors_construction}. We use four combinations of factors with one, three, five, and seven factors. They are denoted by 1F (market), 3F (market, size, and value), 5F (3F + gross profitability and investment), and 7F (5F + asset growth and accruals), respectively.

\subsection{Sector Classification}\label{subsection_sector_classification}

Each stock is classified into one of 10 sectors, following the Standard Industrial Classification (SIC). Table \ref{table_sic} shows the number of companies from our sample in each sector. Note that some groups are quite large (the group Others, for instance, has more than 100 stocks). This motivates the use of additional restrictions we discussed in Section \ref{subsection_sigma_epsilon}.

\begin{table}[H]
\caption{\textbf{Number of Stocks per Sector}}
\label{table_sic}
\centering
\begin{minipage}{\linewidth}
\begin{footnotesize}
The table reports the number of stocks in each of the ten sectors considered. The sector classification follows the Standard Industrial Classification (SIC).
\end{footnotesize}
\end{minipage}
\begin{threeparttable}
\begin{tabular}{lc}
\hline
\Tstrut
Sector                                     & Number of Stocks  \\
\hline
\Tstrut
Consumer Non-Durables                      & 31 \\
Consumer Durables                          & 8  \\
Manufacturing                              & 65 \\
Oil, Gas, and Coal Extraction              & 32 \\
Business Equipment                         & 61 \\
Telecommunications                         & 10 \\
Wholesale and Retail                       & 45 \\
Health Care, Medical Equipments, and Drugs & 26 \\
Utilities                                  & 36 \\
Others                                     & 116 \\
\hline
\end{tabular}
\end{threeparttable}
\end{table}

\subsection{Residual Covariance Matrices}
\label{subsection_residual_covariance}

\begin{figure}[H]
\caption{\textbf{Realized Raw Correlations}}
\begin{footnotesize}
\begin{spacing}{}
The blue dots represent the correlations between stocks larger than 0.15 in absolute value in at least 1/3 of the sample days. Red squares represent the groups defined by sector industrial classification (SIC). The axes have indices that correspond to the 430 stocks in our sample. The correlations are computed using the sequence of realized covariance matrices before any factor decomposition.
\bigskip
\end{spacing}
\end{footnotesize}
\centering
\includegraphics[width = 0.79\textwidth]{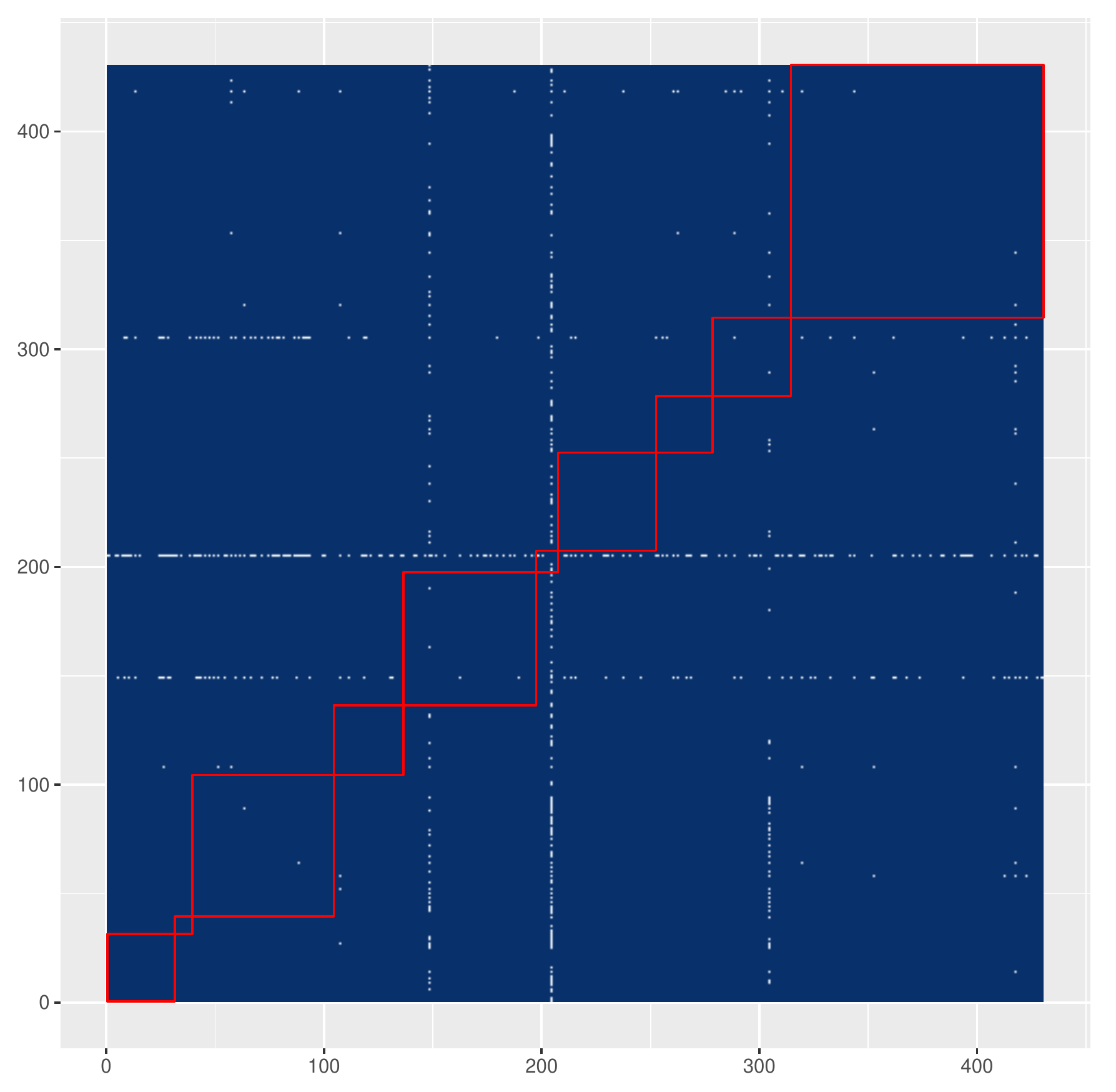}
\label{fig_block_diagonal}
\end{figure}

We analyze the residual covariance matrices for the four different combinations of factors. We use the previous notation, $\bs\Sigma_{\epsilon,t}= \bs\Sigma_{t}-\bs B_{t}'\bs\Sigma_{f,t}\bs B_{t}$, for each of the 1,495 days in the sample. To analyze whether these matrices are approximately block diagonal, we follow the procedure discussed in \citet{ait2016using}. First, we transform the covariance matrices into correlation matrices by dividing the entries by the standard deviations. We classify a correlation as significant if it is higher than 0.15 in absolute value in at least 1/3 of the sample. We then plot the significant relations as dots while the remaining points are blank. Sectors in Table \ref{table_sic} are represented as red squares (in the same order). Figure \ref{fig_block_diagonal} shows the results when we apply the criterion to the full covariance matrices $\bs\Sigma_{t}$ before the factor decomposition. Most correlations are classified as significant.

\begin{figure}[H]
\caption{\textbf{Realized Residual Correlations}}
\begin{footnotesize}
\begin{spacing}{}
The blue dots represent the correlations between stocks larger than 0.15 in absolute value in at least 1/3 of the sample days. Red squares represent the groups defined by sector industrial classification (SIC). The axes have indices that correspond to the 430 stocks in our sample. This plot shows the results for four series of residual covariance matrices obtained after using four different factor decompositions.
\bigskip
\end{spacing}
\end{footnotesize}
\includegraphics[width = \textwidth]{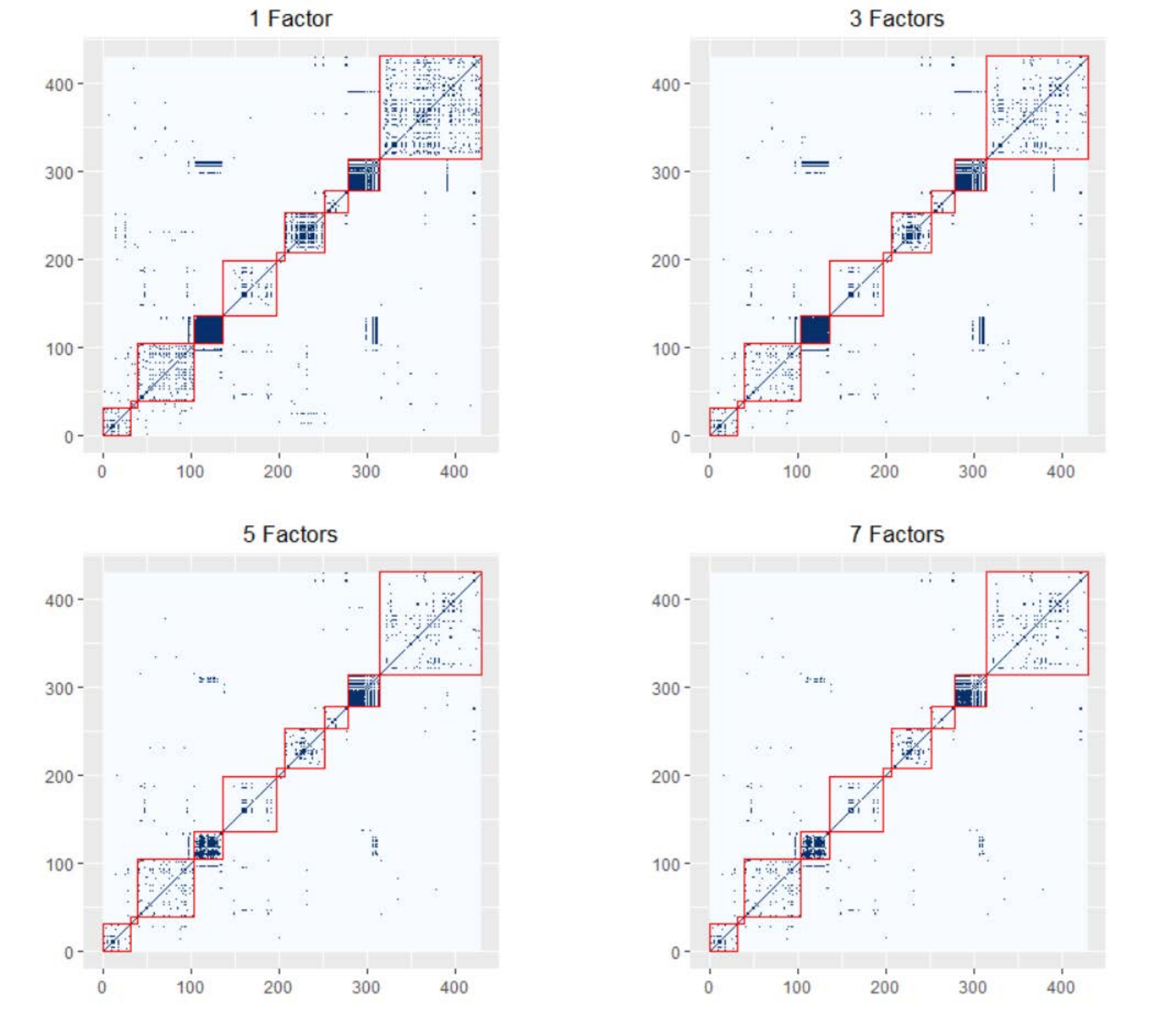}
\label{fig_block_diagonal_factors}
\end{figure}

When we follow this procedure for the residual covariance matrices (Figure \ref{fig_block_diagonal_factors}), we obtain plots that are much more sparse than previously. This means that most correlations are not significant according to the criterion. Furthermore, we can see that most of the dots are contained inside the blocks, meaning that the most significant correlations are between stocks in the same sector. Therefore, we interpret the three plots as evidence that we would not lose much information by assuming block diagonality for the residual covariance matrices. Note that the results are robust across different factor configurations.

\section{Forecasting Results}\label{S:Results}

This section reports the forecasting results obtained using the methodology described in Section \ref{section_forecasting_methodology}. Subsection \ref{subsection_results_factors} shows the forecasting results for $\bs\Sigma_{f}$, while Subsection \ref{subsection_results_complete} shows the results for the complete covariance matrix $\bs\Sigma$. Daily forecasts are computed using rolling windows of 1,000 observations in both cases. Since we need 22 days to compute the first monthly regressor, $\bs\Sigma_{f,t}^{month}$, the out-of-sample forecasts comprise days 1,023 to 1,495 ($T_1$ to $T_2$), totaling 473 daily forecasts.

We evaluate our forecasts by using the $\ell_2$-norm for the vector of errors, that is,  $||\widehat{\bs e}_{T+1}||=||\textnormal{vech}(\widehat{\bs\Sigma}_{T+1}-\bs\Sigma_{T+1})||$. We compare different methods by using the average $\ell_2$-forecast error:
\begin{equation}
\textnormal{average $\ell_2$-forecast error}=\frac{1}{T_2-T_1+1}\sum_{T=T_1}^{T=T_2}||\widehat{\bs\epsilon}_{T+1}||.
\end{equation}
In all cases, our reference forecast is a random walk model: $\widehat{\bs\Sigma}_{f, T+1}=\bs\Sigma_{f,T}$ and $\widehat{\bs\Sigma}_{T+1}=\bs\Sigma_{T}$ for the complete covariance matrix.

\subsection{Factor Covariance Matrix}\label{subsection_results_factors}

Table \ref{table_factors_covariance_errors} shows the forecast results for the factor covariance matrices, following the method described in Section \ref{forecasting_factors_covariance} (we refer to this method as FHAR, from factor HAR, hereafter). However, the results are uniformly stronger when we apply a log-matrix transformation, as proposed by \citet{chiu1996matrix}. Before estimation, we apply the log-matrix transformation to all data, that is, $\bs\Omega_{f,t}=\log(\bs\Sigma_{f,t})$. We then use the FHAR model to compute $\widehat{\bs\Omega}_{f,t+1}$. Finally, we revert the transformation by applying the exponential-matrix transformation and obtain our forecasts, that is, $\widehat{\bs\Sigma}_{f,t+1}=\exp(\widehat{\bs\Omega}_{f,t+1})$. We also report the results without applying the log-matrix transformation.

\begin{table}[H]
\caption{\textbf{Forecast Precision for Factor Covariance Matrices}}
\label{table_factors_covariance_errors}
\centering
\begin{minipage}{\linewidth}
\begin{footnotesize}
$\ell_2$ represents the average $\ell_2$-forecast error over the 473 out-of-sample days, that is,
\[
\textnormal{average $\ell_2$-forecast error}=\frac{1}{T_2-T_1+1}\sum_{T=T_1}^{T=T_2}||\widehat{\bs\epsilon}_{T+1}||.
\]
$\ell_2/\ell_{2,RW}$ represents the ratio of the $\ell_2$-forecast error for other methods to the random walk value. FHAR is the factor heterogeneous autoregressive model described in Section \ref{forecasting_factors_covariance}. Each line represents a different factor configuration. Numbers shown between parenthesis represent the results obtained with the adaLASSO method.
\end{footnotesize}
\end{minipage}
\begin{threeparttable}
\bigskip
\begin{tabular}{cccccc}
\hline
&  & $\ell_2$ & {\ul } & \multicolumn{2}{c}{$\ell_2$ / $\ell_{2,RW}$} \\
\cline{3-3}
\cline{5-6}
Model &  & Random Walk  &        & FHAR           & FHAR, Log-matrix     \\
        \hline
        1F    &  & 0.40  &        & 0.96 (0.96)     & 0.92 (0.92)          \\
        3F    &  & 0.44  &        & 0.98 (0.97)     & 0.90 (0.90)        \\
        5F    &  & 0.51  &        & 0.95 (0.95)     & 0.89 (0.89)          \\
        7F    &  & 0.62  &        & 0.99 (1.04)     & 0.86 (0.87)        \\
\hline
\end{tabular}
\end{threeparttable}
\end{table}

From Table \ref{table_factors_covariance_errors}, we see that using the log-matrix transformation considerably improves the forecasts. After this transformation, the variance and covariance series become smoother than in the original data. This reduces the weights of outliers when fitting the model and improves forecasting precision. Another advantage is that the exponential matrix is positive definite by construction. This consideration is important, as we will need the inverse of the covariance matrix $\bs\Sigma$ when solving the minimum variance problems in Section \ref{sec_portfolio}. Finally, applying adaLASSO to the best-performing models does not improve the results up to the second decimal place.

\begin{figure}[H]
\caption{\textbf{Average Number of Selected Variables in FHAR Log-matrix LASSO models}}
\begin{footnotesize}
\begin{spacing}{}
These panels show the daily average number of variables selected by the LASSO method for variance and covariance equations on a given day. We sort the panels by the number of factors in the model (increasing from top to bottom) and by variable class (variances on the left and covariances on the right). Blue lines are local polynomial regressions. The maximum numbers of predictive variables for each configuration are 18 (3 factors), 45 (5 factors), and 76 (7 factors).
\bigskip
\end{spacing}
\end{footnotesize}
\centering
\includegraphics[width = \textwidth]{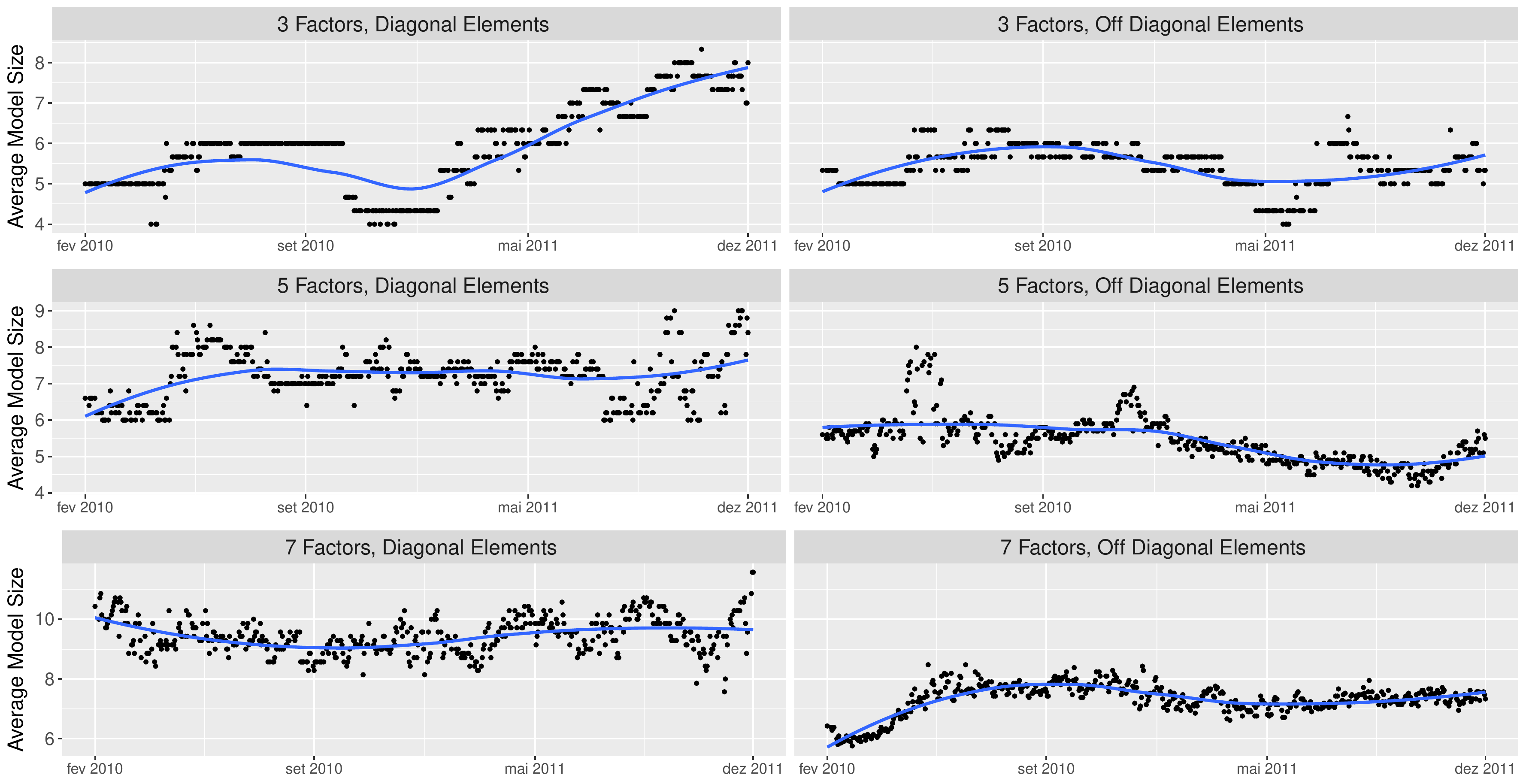}
\label{fig_parameter_size}
\end{figure}

Since we estimate all equations daily, we now investigate their evolution over time\footnote{The results for average equation size and change in the number of selected variables obtained with the adaLASSO are similar to the ones obtained via LASSO. These results are readily upon request.}. Figure \ref{fig_parameter_size} shows the average number of variables selected by LASSO over the 473 days for the FHAR Log-matrix models. To illustrate, consider the upper-left panel in this figure. It shows the average model size for the diagonal equations in the FHAR Log-matrix model with three factors. In other words, it displays the mean size for the equations of $\sigma_{market}^2$, $\sigma_{SMB}^2$, and $\sigma_{HML}^2$. Panels on the right show the same results for the covariance equations. From top to bottom, we vary the number of factors\footnote{For the 1F configuration, we forecast only $\sigma_{market}^2$. In this case, the equations estimated by LASSO and adaLASSO have only three predictors, which are, in this case, always selected. For this reason, we do not present the average number of selected variables (constant) or average change in the number of selected variables (zero).}.

In all cases, LASSO reduces the number of selected variables substantially. This reduction is most noticeable for the FHAR Log-matrix model with seven factors, in which the number of selected variables fluctuates around 10 (out of 76 potential predictors) for the variance equations and even less for the covariance equations.

\begin{figure}[H]
\caption{\textbf{Average Change in the Number of Selected Variables in FHAR Log-matrix models}}
\begin{footnotesize}
\begin{spacing}{}
These panels show the daily average change in the number of variables selected by LASSO for variance and covariance equations. We count a change as when a variable goes from zero to non-zero or from non-zero to zero. The results are a percentage relative to each equation's maximum number of variables. We sort the panels by the number of factors in the model (increasing from top to bottom) and by variable class (variances on the left and covariances on the right).  Blue lines are local polynomial regressions.
\bigskip
\end{spacing}
\end{footnotesize}
\centering
\includegraphics[width = \textwidth]{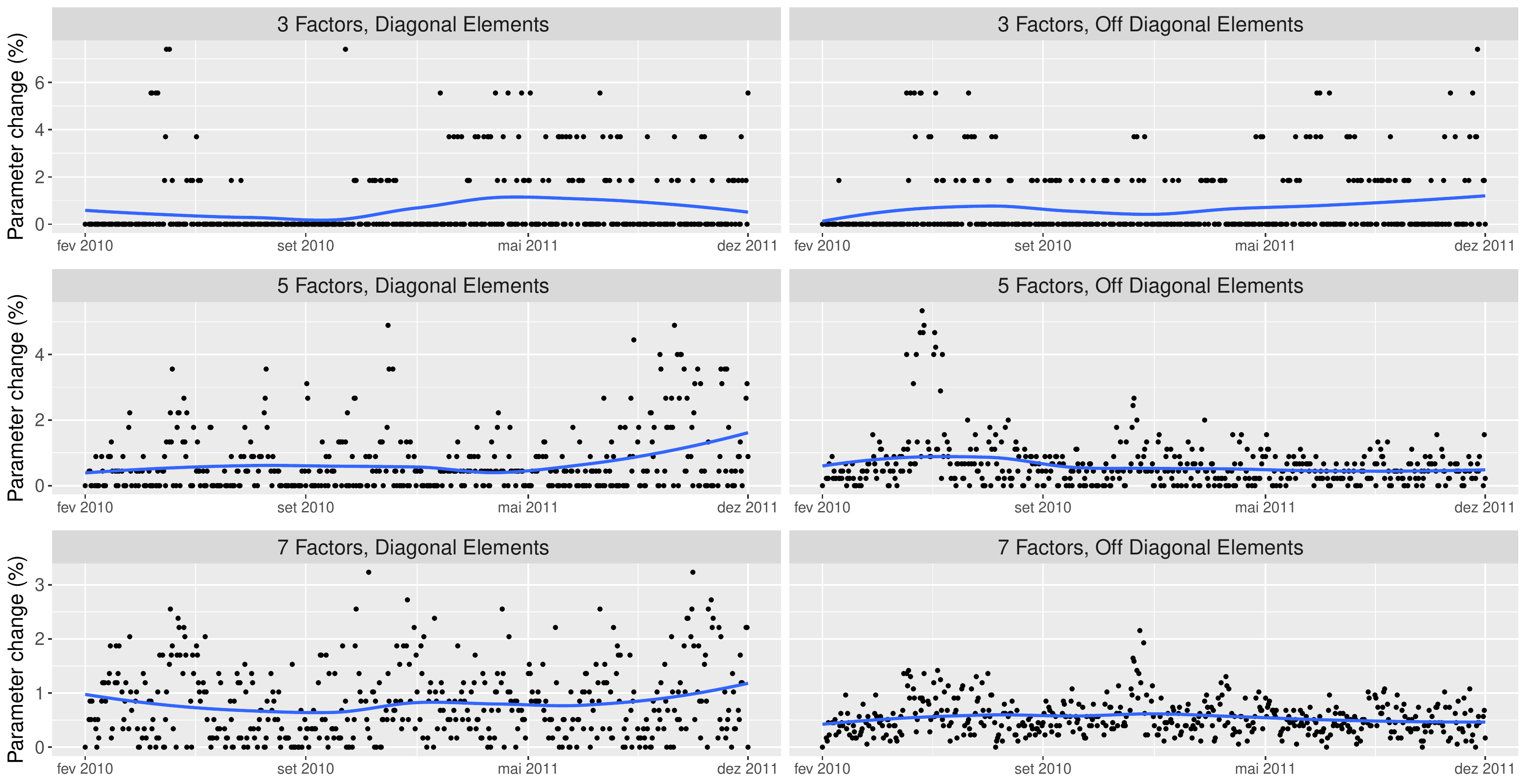}
\label{fig_parameter_change}
\end{figure}

Another feature is that average equation sizes are stable over time. Figure \ref{fig_parameter_change} shows the average change in the number of selected variables from one day to the next. We consider parameters changing from zero to non-zero values or the opposite direction. All results are reported in percentage (relative to the maximum number of variables in each equation). For the three-factor models, variance and covariance equations remain unchanged on most days. Despite the stronger variation in the five- and seven-factor configurations, the percentage changes are smaller than 5\% on almost all days. In general, variance equations are slightly more stable. We present results for the residual covariance matrix forecasts in Appendix \ref{appendix_residual_parameter_selection}.

\subsection{Complete Covariance Matrix}\label{subsection_results_complete}

Table \ref{table_complete_covariance_errors} shows the forecast results for the complete covariance matrix (430 stocks). In this case, VHAR denotes the complete methodology described in Section \ref{section_forecasting_methodology}. We compute all models using Log-matrix transformation and estimate these models using LASSO and adaLASSO. Table \ref{table_complete_covariance_errors} also reports alternative models for comparison purposes:
\begin{enumerate}
\item
Random walk (RW): the standard reference model throughout this work.
\item
Exponentially weighted moving average (EWMA) with smoothing parameter $\lambda=0.96$, as recommended by the RiskMetrics methodology.
\item
The dynamic conditional correlation, BEKK, and approximate factor models, all with nonlinear shrinkage, as proposed by \citet{engle2017large} and  \citet{DeNard2021}. We apply the DCC-NL model with the market factor for the approximate factor model, as recommended in \citet{DeNard2021}. These approaches will be called DCC-NL, BEKK-NL, and AFM1-DCC-NL, hereafter. We also consider the IDR-DCC-NL model proposed by \citet{deNard2022}.
\item
Modified versions of the random walk model (Blocks 1F, 3F, 5F, and 7F). In these cases, we decompose the original realized covariance matrices using factors and impose block diagonality on the residual covariance matrices. Instead of using VHAR models to forecast future realizations, we use the random walk model with the adjusted covariance matrices as forecasts. By doing this, we try to approximate the methodology used by \citet{fan2016factors}\footnote{\citet{fan2016factors} impose block diagonality when using the high-frequency data for factors and returns to estimate the integrated covariance. The authors use the GICS system to classify stocks in sectors. The model used for forecasting and portfolio allocation is the random walk.}.
\end{enumerate}

Some alternative models use daily returns data instead of realized covariance matrices as inputs. We compare our results to these models to see whether high-frequency data bring gains in portfolio performance.

\begin{table}[H]
\caption{\textbf{Forecast Precision for Complete Covariance Matrices}}
\label{table_complete_covariance_errors}
\centering
\begin{minipage}{\linewidth}
\begin{footnotesize}
$\ell_2$ represents the average $\ell_2$-forecast error, computed for 473 days, that is,
\[
\textnormal{average $\ell_2$-forecast error}=\frac{1}{T_2-T_1+1}\sum_{T=T_1}^{T=T_2}||\widehat{\bs\epsilon}_{T+1}||.
\]
$\ell_{2,RW}$ is the error for the random walk model. $\ell_2/\ell_{2,RW}$ represents the ratio between the error values. The column ``Latent Covariance Models'' displays the results for models based on daily return data. The columns ``VHAR'' and ``VHAR, Log-matrix'' show our results (without and with log-matrix transformation, respectively). 1F, 3F, 5F, and 7F stand for the 1-factor, 3-factor, 5-factor, and 7-factor configurations for the factor covariance matrix.
\end{footnotesize}
\end{minipage}
\begin{threeparttable}
\bigskip
\begin{tabular}{llll} \hline  \Tstrut
Model (Benchmarks) & $\ell_2/\ell_{2,RW}$     & VHAR (Log-matrix) & $\ell_2/\ell_{2,RW}$    \Bstrut     \\  \hline
EWMA (Returns)                & 6.93      & 1F, LASSO      & 0.86 \\
BEKK-NL            &1.72      & 3F, LASSO         & 0.85 \\
DCC-NL             & 1.72     & 5F, LASSO         & 0.87 \\
AFM1-DCC-NL        & 1.72     & 7F, LASSO         & 0.86 \\
IDR-DCC-NL         & 1.72     & 1F, adaLASSO      & 0.86 \\
Block 1F           & 0.98     & 3F, adaLASSO      & 0.86 \\
Block 3F           & 0.97     & 5F, adaLASSO      & 0.86 \\
Block 5F           & 0.98     & 7F, adaLASSO      & 0.86 \\
Block 7F           & 0.98
\Bstrut   \\  \hline \Tstrut
Random Walk (RW) $\ell_{2,RW}$    & 341.57 &                   &     \Bstrut  \\ \hline
\end{tabular}
\end{threeparttable}
\end{table}

Models based on daily returns do not outperform the benchmark, with a particularly bad performance for the EWMA. Despite having performed much better, BEKK-NL, DCC-NL, AFM1-DCC-NL, and IDR-DCC-NL models have almost two times larger errors than the random walk model. Imposing block diagonality on the residual covariance matrices (Blocks 1F, 3F, 5F, and 7F) improves the results, relative to the random walk, by 3\%. Our methodology can beat the random walk model by up to 15\%. In terms of the average $\ell_2$-forecast error, there is no difference when estimating the models with LASSO or adaLASSO.

\section{Portfolio Selection}\label{sec_portfolio}
Despite the encouraging results obtained when forecasting covariance matrices, one could argue that our measure of forecast precision is too aggregate, especially considering that we have more than 90,000 unique entries being forecasted simultaneously. For instance, these estimates may produce large mistakes in only a few entries but with large implications when applied to practical problems. To evaluate the economic benefits provided by our forecasts, we use them to construct daily investment portfolios\footnote{In this section, we use daily covariance matrix forecasts to build portfolios for the next day. These portfolios are rebalanced daily, using the most recent forecasts. We investigate portfolios formed for longer holding periods in Appendix \ref{partialrebalancing}. Similar results to this section hold even with infrequent rebalancing.}. We construct minimum variance portfolios using the models described in the previous section. We consider the global minimum variance portfolio (without any restrictions on weights), the restricted portfolio (limiting long positions and short-selling), and the long-only portfolios (without any short-selling).

We focus our analysis on variations of the minimum variance problem, as the sole portfolio problem requires only information on the covariance matrix. Other portfolio optimization problems, such as finding tangency portfolios that maximize Sharpe ratios, also require estimates of expected returns for every eligible asset. When we analyze the outcomes, we also focus on risk measures, such as the portfolio standard deviation, as we aim to minimize the portfolio's risk. In finance theory, all efficient portfolios have higher expected returns and Sharpe ratios than the minimum variance portfolio. Minimum variance optimizations that are less precise or biased may, by chance, find portfolios with higher realized returns and higher Sharpe ratios. While we should base our assessment of the models on the risk measures, other measures related to the concentration, trading volume, and even realized return might also provide relevant information.

First, we compute various statistics for indices directly related to our portfolio exercise. Table \ref{table_sp_indexes} shows the standard deviation, kurtosis, skewness, average excess return over the risk-free rate, and Sharpe ratio for the daily series of returns for three S\&P 500 indices: minimum volatility, low volatility, and equal weight. We provide these values as reference points to evaluate our results, as these indices are commonly used by practitioners. According to the index provider, the first index ``is designed to reflect a managed-volatility equity strategy that seeks to achieve lower total risk, measured by standard deviation, than the S\&P 500 while maintaining similar characteristics.'' It is based on the Northfield Open Optimizer and is rebalanced every six months. The second index is based on risk measures for each asset individually, focusing on stocks with low volatility.

\begin{table}[H]
\caption{\textbf{S\&P Index Performances}}
\label{table_sp_indexes}
\centering
\begin{minipage}{\linewidth}
\begin{footnotesize}
Average excess returns are defined as the annualized average excess returns over the risk-free rate for the index total return series. Standard deviations are calculated for these series and are also annualized. The Sharpe ratio is computed using these two quantities. The sample is from 2010/02/12 to 2011/12/29, which corresponds to 473 days considered in the portfolio exercises.
\end{footnotesize}
\end{minipage}
\begin{threeparttable}
\bigskip
\begin{tabular}{lcccc} \hline  \Tstrut
& \begin{tabular}[c]{@{}c@{}}S\&P 500 \end{tabular}
& \begin{tabular}[c]{@{}c@{}}S\&P 500 \\ Min. Vol.\end{tabular}
& \begin{tabular}[c]{@{}c@{}}S\&P 500 \\ Low Vol.\end{tabular}
& \begin{tabular}[c]{@{}c@{}}S\&P 500 \\ Equal Weight\end{tabular}
\Bstrut
\\ \hline
\Tstrut
Standard Deviation (\%)                 & 21.00     & 16.41     & 14.84    & 23.48   \\
Kurtosis                                & 2.88      & 3.37      & 3.54     & 2.67    \\
Skewness                                & --0.35    & --0.36    & --0.38   & --0.34   \\
Average Excess Return (\%)              & 10.51     & 13.00     & 13.50    & 12.78   \\
Cumulative Return (\%)                  & 17.11     & 24.70     & 26.47    & 20.93   \\
Sharpe Ratio                            & 0.50      & 0.79      & 0.91     & 0.54    \Bstrut
\\ \hline
\end{tabular}
\end{threeparttable}
\end{table}

\subsection{Global Minimum Variance Portfolios}\label{global_minimum_variance}
Consider the problem of an investor at time $t=t_0,\ldots,T-1$ who wishes to construct a minimum variance portfolio to be held in time $t+1$. For this minimization problem, the investor needs to forecast the future covariance matrix, $\hat{\Sigma}_{t+1}$. The optimization problem consists of choosing a vector of weights $\hat{w}_{t+1}$ (dimension $N \times 1$):
\begin{equation}\label{eq_minimum_variance}
\begin{gathered}
\widehat{\bs w}_{t+1}= \underset{\bs w_{t+1}}{\textnormal{arg min}} \quad \bs w_{t+1}'\widehat{\bs\Sigma}_{t+1}\bs w_{t+1} \\
\textnormal{subject to} \quad \bs w_{t+1}'\bm{1}=1.
\end{gathered}
\end{equation}

We use our VHAR method with log matrix transformation, as it provides the best results in terms of forecasting precision. Estimation is done via LASSO and adaLASSO. We compare our results against the random walk models (RW and Block models), the EWMA, BEKK-NL, DCC-NL, AFM1-DCC-NL, and IDR-DCC-NL models. We evaluate ex-post portfolio performance, using our time $t$ estimated weights, $\widehat{\bs w}_{t+1}$, with data from $t+1$. In the following, $\widehat{w}_{it}$ is the $i$-th component of $\widehat{\bs w}_t$. We also define $r_{pt}^*=(r_{pt}-\frac{1}{(T-t_0)}\sum_{t=t_0+1}^{T}r_{pt})$.
We show the following statistics:
\begin{enumerate}
\item
Standard deviation: $\sigma_p=\sqrt{ \frac{1}{(T-t_0)} \sum_{t_0+1}^{T}(r_{pt}^*)^2}$.
\item
Lower partial standard deviation: $\sqrt{ \frac{1}{\sum_{t_0+1}^{T}I(r_{pt}^*<0)} \sum_{t_0+1}^{T}(r_{pt}^*)^2 * I(r_{pt}^*<0)}$.
\item
Kurtosis: $\frac{\sum_{t=t_0+1}^{T}(r_{pt}^*)^{4}/(T-t_0)}{\sigma_p^{4}}  - 3$.
\item
Skewness: $\frac{\sqrt{(T-t_0)(T-t_0-1)}}{T-t_0-2}\frac{\sum_{t=t_0+1}^{T}(r_{pt}^*)^{3}/(T-t_0)} {\sigma_p^{3}}$.
\item
Average diversification ratio: $\frac{1}{(T-t_0)} \sum_{t_0+1}^{T} \frac{\sum_{i=1}^{N}\widehat{w}_{it}\sigma_{it}}{\sigma_{pt}} $, where $\sigma_{pt}=\widehat{\bs w}_{t}'\bs\Sigma_{t}\widehat{\bs w}_{t}$.
\item
Average max. weight: $\frac{1}{(T-t_{0})}\sum_{t=t_0+1}^{T}\max_{1 \leq i \leq N}(\widehat{w}_{it})$ for  $i=1,\ldots,N$.
\item
Average min. weight: $\frac{1}{(T-t_{0})}\sum_{t=t_0+1}^{T}\min_{1 \leq i \leq N}(\widehat{w}_{it})$ for  $i=1,\ldots,N$.
\item
Average gross leverage: $\frac{1}{N(T-t_{0})}\sum_{t=t_0+1}^{T}\sum_{i=1}^{N}|\widehat{w}_{it}|$.
\item
Proportion of leverage: $\frac{1}{N(T-t_{0})}\sum_{t=t_0+1}^{T}\sum_{i=1}^{N}I(\hat{w}_{it}<0)$.
\item
Average turnover: $\frac{1}{N(T-t_{0})}\sum_{t=t_0+1}^{T}\sum_{i=1}^{N}|\widehat{w}_{it}-\widehat{w}_{it}^{hold}|$, where $\widehat{w}_{it}^{hold}=\widehat{w}_{it-1}\frac{1+r_{it-1}}{1+r_{pt-1}}$. $r_{pt}$ is the portfolio return at time $t$, $r_{it}$ is the stock $i$ return at time $t$, and $\widehat{w}_{it}^{hold}$ is the weight of stock $i$ in the hold portfolio. The hold portfolio at time $t+1$ is defined as the resulting portfolio from keeping the number of shares fixed during period $t$.
\item Average excess return: $\mu_{p}^e=\frac{1}{(T-t_0)}\sum_{t=t_0+1}^{T}(r_{pt}^e)=\frac{1}{(T-t_0)}\sum_{t=t_0+1}^{T}(\widehat{\bs w}_{t}'\bs r_t-r_{f,t})$, where $r_{f,t}$ is the risk-free rate.
\item
Cumulative Return: $\prod_{t=t_0+1}^{T}(1+r_{pt})$.
\item
Sharpe ratio: $\frac{\mu_p^e}{\sigma_p}$.
\end{enumerate}

Table \ref{table_minimum_variance} presents the minimum variance portfolio results. Regarding standard deviations, the VHAR and Block RW models perform better than all others. For these models, the standard deviations go from a high of 8.46 \% (one factor and LASSO) to a low of 8.09 \% (seven factors and adaLASSO). The lowest value is almost half of the standard deviation for the S\&P 500 Minimum Volatility Index in Table \ref{table_sp_indexes}. Benchmark models (upper results) also have a smaller standard deviation than the S\&P 500 indices (except for the EWMA, which has the worst performance in this sense). Similar results hold for lower partial standard deviations. Among the models that do not use intraday data, the AFM1-DCC-NL model has superior performance.

Some portfolios have extreme short positions. In most cases, the leverage proportion is close to 50\%. Average gross leverage is also quite high for many of the alternative models. Since shorting stocks may not be feasible and can be costly, we consider these positions potentially extreme. The values for average maximum and minimum weights are also quite large. Our models appear to generate more balanced portfolios.

Our models are slightly inferior to the IDR-DCC-NL model in terms of average excess and cumulative returns. The VHAR with five factors and LASSO estimation is the one with close performance to the IDR-DCC-NL alternative. Our models have superior performance compared to the RW and Block models and are much better than the EWMA specification. Since the optimization problem focuses on minimizing the portfolio variance, it is not obvious which models should achieve higher returns. For this reason, we do not see the slightly superior performance of the IRD-DCC-NL model as a problem. Moreover, the small difference in realized returns is more than compensated by the substantial reduction in standard deviation (as evidenced by the higher Sharpe ratios for our models).

\begin{landscape}
\begin{table}[H]
\caption{\textbf{Statistics for Daily Portfolios - Global Minimum Variance}}
\label{table_minimum_variance}
\begin{minipage}{\linewidth}
\begin{footnotesize}
The table reports the results for portfolios constructed according to the optimization problem as in (\ref{eq_minimum_variance}). Different models are used to provide the one-step-ahead forecasts for the realized covariance matrix: RW is the random walk model, and Blocks 1F, 3F, 5F, and 7F are random walks applied to the residual covariance matrix after the four different factor decompositions.
\end{footnotesize}
\end{minipage}
\begin{threeparttable}
\bigskip
\resizebox{\linewidth}{!}{
\begin{tabular}{ccccccccccc}
\hline
& RW & Block 1F & Block 3F & Block 5F & Block 7F & EWMA (Returns)& BEKK-NL & DCC-NL &  AFM1-DCC-NL & IDR-DCC-NL \\
                                      \hline
Standard Deviation (\%)               & 12.07   & 8.21    & 8.29    & 8.25    & 8.25    & 14.62  & 9.41   & 10.65   &  9.24  & 10.36 \\
Lower Partial Standard Deviation (\%) & 12.82   & 8.79    & 8.94    & 8.73    & 8.83    & 14.90  & 9.63   & 11.31   & 9.40   & 11.09 \\
Kurtosis                              & 1.29    & 0.69    & 0.77    & 0.72    & 0.73    & 2.12   & 1.60   & 4.60    & 1.69   & 3.86  \\
Skewness                              & --0.35  & --0.39  & --0.40  & --0.38  & --0.40  & --0.15 & --0.33 & -0.50   & -0.05  & -0.56 \\
Average Diversification Ratio         & 3.29    & 6.05    & 6.11    & 6.11    & 6.12    & 1.01   & 3.03   & 3.53    & 3.15   & 3.49  \\
Average Max. Weight                   & 0.06    & 0.10    & 0.10    & 0.10    & 0.10    & 0.20   & 0.06   & 0.12    & 0.13   & 0.15  \\
Average Min. Weight                   & --0.06  & --0.03  & --0.03  & --0.03  & --0.03  & --0.15 & --0.04 & --0.04  & --0.05 & --0.03 \\
Average Gross Leverage                & 5.94    & 3.08    & 3.14    & 3.14    & 3.19    & 12.55  & 5.09   & 4.11    & 4.47   & 3.65  \\
Proportion of Leverage (\%)           & 44.30   & 44.40   & 44.22   & 44.10   & 44.11   & 49.17  & 45.11  & 51.73   & 46.91  & 50.5  \\
Average Turnover (\%)                 & 1.80    & 0.75    & 0.78    & 0.78    & 0.80    & 0.27   & 0.11   & 0.21    & 0.16   & 0.35  \\
Average Excess Return (\%)            & 14.20   & 12.72   & 14.46   & 15.37   & 14.95   & 3.42   & 17.98  & 17.46   & 16.16  & 19.21 \\
Cumulative Return (\%)                & 29.04   & 26.42   & 30.59   & 32.86   & 31.82   & 4.74   & 39.27  & 37.58   & 34.63  & 42.25 \\
Sharpe Ratio                          & 1.18    & 1.55    & 1.74    & 1.86    & 1.81    & 0.23   & 1.91   & 1.64    & 1.75   & 1.85  \\
\hline
& \multicolumn{2}{c}{1 Factor}  &
\multicolumn{2}{c}{3 Factors}   &
\multicolumn{2}{c}{5 Factors}   &
\multicolumn{2}{c}{7 Factors}\\
& \multicolumn{2}{c}{VHAR}      &
\multicolumn{2}{c}{VHAR}        &
\multicolumn{2}{c}{VHAR}        &
\multicolumn{2}{c}{VHAR}       \\
& \multicolumn{2}{c}{(Log matrix)} &
\multicolumn{2}{c}{(Log matrix)} &
\multicolumn{2}{c}{(Log matrix)} &
\multicolumn{2}{c}{(Log matrix)} \\
& LASSO             & adaLASSO          & LASSO             & adaLASSO     &  LASSO             & adaLASSO          & LASSO             & adaLASSO \\
\hline
Standard Deviation (\%)               & 8.46   & 8.42   & 8.37   & 8.32    &  8.29 & 8.25   & 8.12   & 8.09   \\
Lower Partial Standard Deviation (\%) & 8.86   & 8.81   & 8.78   & 8.68    &  8.57 & 8.53   & 8.52   & 8.51   \\
Kurtosis                              & 0.96   & 0.98   & 0.99   & 0.98    &  0.96 & 0.94   & 1.13   & 1.06   \\
Skewness                              & --0.21  & --0.20  & --0.18  & --0.17   & --0.15 & --0.13  & --0.24  & --0.22  \\
Average Diversification Ratio         & 4.79   & 4.81   & 5.02   & 5.03    &  4.87 & 4.88   & 4.96   & 4.96   \\
Average Max. Weight                   & 0.07   & 0.08   & 0.08   & 0.08    &  0.08 & 0.09   & 0.08   & 0.09   \\
Average Min. Weight                   & --0.02  & --0.02  & --0.02  & --0.02   & --0.02 & --0.02  & --0.02  & --0.03  \\
Average Gross Leverage                & 2.66   & 2.67   & 2.80   & 2.80    &  2.82 & 2.82   & 2.93   & 2.93   \\
Proportion of Leverage (\%)           & 45.89  & 46.01  & 44.88  & 45.03   & 44.89 & 45.12  & 45.26  & 45.50  \\
Average Turnover (\%)                 & 0.20   & 0.22   & 0.20   & 0.22    &  0.19 & 0.21   & 0.20   & 0.22   \\
Average Excess Return (\%)            & 15.24  & 15.18  & 17.69  & 17.45   & 18.93 & 18.61  & 18.09  & 17.85  \\
Cumulative Return (\%)                & 32.49  & 32.35  & 38.74  & 38.13   & 42.01 & 41.19  & 39.85  & 39.21  \\
Sharpe Ratio                          & 1.80   & 1.80   & 2.11   & 2.10    &  2.28 & 2.26   & 2.23   & 2.21   \\
\hline
\end{tabular}}
\end{threeparttable}
\end{table}
\end{landscape}

\subsection{Restricted Minimum Variance Portfolios}\label{restricted_minimum_variance_portfolio}
We now resort to what we consider a more realistic investor problem. In this section, we solve a problem similar to equation (\ref{eq_minimum_variance}), except that now we impose two additional restrictions. First, we allow the maximum leverage to be 30\% (in some sense, consistent with a 130-30 fund concept in the mutual fund industry). Second, we restrict the maximum weights on individual stocks to be 20\% (in absolute value). The problem for an investor at time $t=t_0,\ldots, T-1$ is then given by

\begin{equation}\label{eq_minimum_variance_restricted}
\begin{gathered}
\widehat{\bs w}_{t+1}= \underset{\bs w_{t+1}}{\textnormal{arg min}} \quad \bs w_{t+1}'\widehat{\bs\Sigma}_{t+1}w_{t+1} \\
\textnormal{subject to} \quad \bs w_{t+1}'\bm{1}=1, \\
\sum_{i=1}^{N}|w_{it+1}|I(w_{it}<0)\leq 0.30 \quad \textnormal{and} \quad |w_{it+1}|\leq0.20.
\end{gathered}
\end{equation}

In Table \ref{table_restricted_portfolio}, we report the same performance statistics as in the last subsection. Again, our methodology generates lower standard deviations than the competing models. The VHAR with five factors and estimated with LASSO has the smallest standard deviation (12.57). In terms of returns, the Block 5F model is the best-performing alternative. This is the best model also in terms of Sharpe Ratios. The several VHAR specifications come second. One additional interesting result is that the IDR-DCC-NL model now performs better than the DCC-NL and AFM1-DCC-NL models, both in terms of standard deviation as cumulative and average returns.

\begin{landscape}
\begin{table}[H]
\caption{\textbf{Statistics for Daily Portfolios - Restricted Minimum Variance}}
\label{table_restricted_portfolio}
\begin{minipage}{\linewidth}
\begin{footnotesize}
The table reports the results for portfolios constructed according to the optimization problem as in (\ref{eq_minimum_variance_restricted}). Different models provide the one-step-ahead forecasts for the realized covariance matrix: RW is the random walk model, and Blocks 1F, 3F, 5F, and 7F are random walks applied to the residual covariance matrix after the four different factor decompositions.
\end{footnotesize}
\end{minipage}
\begin{threeparttable}
\bigskip
\resizebox{\linewidth}{!}{
\begin{tabular}{ccccccccccc}
\hline

                                      & RW & Block 1F & Block 3F & Block 5F & Block 7F & EWMA (Returns)& BEKK-NL & DCC-NL & AFM1-DCC-NL & IDR-DCC-NL \\
                                      \hline
Standard Deviation (\%)               & 13.29  & 13.34  & 13.20  & 13.17   & 13.25  & 15.28  & 15.49  & 14.72  & 16.14 &  13.72 \\
Lower Partial Standard Deviation (\%) & 14.13  & 13.91  & 13.66  & 13.35   & 13.68  & 16.47  & 16.24  & 15.28  & 17.50 &  14.43 \\
Kurtosis                              & 3.40   & 4.71   & 4.68   & 4.79    & 5.10   & 4.16   & 4.36   & 3.08   & 1.66  &  3.66  \\
Skewness                              & --0.27  & --0.13  & --0.08  & --0.10   & --0.17  & --0.48  & --0.36  & --0.15  & --0.35 &  --0.56 \\
Average Diversification Ratio         & 3.55   & 3.92   & 4.11   & 4.12    & 4.07   & 2.22   & 2.24   & 2.15   & 1.64  &  2.51  \\
Average Max. Weight                   & 0.17   & 0.16   & 0.15   & 0.15    & 0.15   & 0.19   & 0.19   & 0.19   & 0.20  &  0.19  \\
Average Min. Weight                   & --0.09  & --0.07  & --0.07  & --0.07   & --0.08  & --0.16  & --0.15  & --0.11  & --0.17 &  --0.06 \\
Average Gross Leverage                & 1.60   & 1.60   & 1.60   & 1.60    & 1.60   & 1.60   & 1.60   & 1.60   & 1.60  &  1.60  \\
Proportion of Leverage (\%)           & 1.91   & 3.11   & 3.08   & 3.06    & 2.93   & 0.71   & 0.85   & 1.41   & 0.58  &  2.44  \\
Average Turnover (\%)                 & 0.43   & 0.40   & 0.42   & 0.41    & 0.42   & 0.09   & 0.10   & 0.11   & 0.03  &  0.19  \\
Average Excess Return (\%)            & 16.72  & 18.23  & 19.01  & 22.42   & 21.22  & 13.68  & 14.24  & 16.91  & 10.04 &  16.91 \\
Cumulative Return (\%)                & 34.88  & 38.74  & 40.83  & 50.14   & 46.79  & 26.74  & 27.99  & 34.86  & 18.06 &  35.22 \\
Sharpe Ratio                          & 1.26   & 1.37   & 1.44   & 1.70    & 1.60   & 0.90   & 0.92   & 1.15   & 0.62  &  1.23  \\
\hline
& \multicolumn{2}{c}{1 Factor} &
\multicolumn{2}{c}{3 Factors}  &
\multicolumn{2}{c}{5 Factors}  &
\multicolumn{2}{c}{7 Factors}  \\
& \multicolumn{2}{c}{VHAR}     &
\multicolumn{2}{c}{VHAR}       &
\multicolumn{2}{c}{VHAR}       &
\multicolumn{2}{c}{VHAR}         \\
& \multicolumn{2}{c}{(Log matrix)} &
\multicolumn{2}{c}{(Log matrix)} &
\multicolumn{2}{c}{(Log matrix)} &
\multicolumn{2}{c}{(Log matrix)} \\
& LASSO             & adaLASSO          & LASSO             & adaLASSO      &   LASSO             & adaLASSO          & LASSO             & adaLASSO \\
\hline
Standard Deviation (\%)               & 13.20   & 13.37  & 12.81  & 12.86 &  12.57  & 12.83  & 12.63  & 12.75  \\
Lower Partial Standard Deviation (\%) & 13.29   & 13.64  & 12.60  & 12.54  &  12.54  & 12.75  & 12.52  & 12.62  \\
Kurtosis                              & 3.89    & 4.09   & 4.82   & 4.84 &  4.44   & 5.05   & 4.47   & 5.14   \\
Skewness                              & --0.03   & --0.12  & 0.14   & 0.12   & 0.04   & 0.10   & 0.03   & 0.05   \\
Average Diversification Ratio         & 3.41    & 3.38   & 3.64   & 3.65   & 3.73   & 3.70   & 3.68   & 3.65   \\
Average Max. Weight                   & 0.15    & 0.15   & 0.15   & 0.15   & 0.15   & 0.15   & 0.15   & 0.15   \\
Average Min. Weight                   & --0.08   & --0.08  & --0.07  & --0.07   & --0.07  & --0.07  & --0.07  & --0.07  \\
Average Gross Leverage                & 1.60    & 1.60   & 1.60   & 1.60    & 1.60   & 1.60   & 1.60   & 1.60   \\
Proportion of Leverage (\%)           & 2.46    & 2.44   & 2.37   & 2.38    & 2.43   & 2.41   & 2.27   & 2.25   \\
Average Turnover (\%)                 & 0.22    & 0.23   & 0.24   & 0.24    & 0.23   & 0.24   & 0.22   & 0.23   \\
Average Excess Return (\%)            & 16.07   & 19.89  & 19.72  & 21.04   & 20.56  & 18.93  & 20.74  & 19.19  \\
Cumulative Return (\%)                & 33.30   & 43.13  & 42.88  & 46.43   & 45.22  & 40.76  & 45.67  & 41.48  \\
Sharpe Ratio                          & 1.22    & 1.49   & 1.54   & 1.64    & 1.64   & 1.48   & 1.64   & 1.51   \\
\hline
\end{tabular}}
\end{threeparttable}
\end{table}
\end{landscape}

\subsection{Long-Only Minimum Variance Portfolios}\label{minimum_variance_portfolio}
In this section, we restrict no short-selling. The problem for an investor at time $t=t_0,\ldots, T-1$ is then given by

\begin{equation}\label{eq_minimum_variance_long_only}
\begin{gathered}
\widehat{\bs w}_{t+1}= \underset{\bs w_{t+1}}{\textnormal{arg min}} \quad \bs w_{t+1}'\widehat{\bs\Sigma}_{t+1}\bs w_{t+1} \\
\textnormal{subject to} \quad \bs w_{t+1}'\bm{1}=1, \\
\quad 0\leq w_{it+1}\leq0.20.
\end{gathered}
 \end{equation}

From Table \ref{table_restricted_portfolio_long_only}, we see that the VHAR family outperforms the benchmarks in almost all cases in terms of standard deviation. The VHAR model with seven factors estimated with LASSO is the best-performing specification.

\begin{landscape}
\begin{table}[H]
\caption{\textbf{Statistics for Daily Portfolios - Restricted Minimum Variance (Long Only)}}
\label{table_restricted_portfolio_long_only}
\begin{minipage}{\linewidth}
\begin{footnotesize}
The table reports the results for portfolios constructed according to the optimization problem as in (\ref{eq_minimum_variance_long_only}). Different models provide the one-step-ahead forecasts for the realized covariance matrix: RW is the random walk model, and Blocks 1F, 3F, 5F, and 7F are random walks applied to the residual covariance matrix after the four different factor decompositions.
\end{footnotesize}
\end{minipage}
\begin{threeparttable}
\bigskip
\resizebox{\linewidth}{!}{
\begin{tabular}{ccccccccccc}
\hline

                                      & RW & Block 1F & Block 3F & Block 5F & Block 7F & EWMA (Returns)& BEKK-NL & DCC-NL & AFM1-DCC-NL & IDR-DCC-NL \\
                                      \hline
Standard Deviation (\%)               & 17.10             & 17.06             & 16.96             & 16.85             & 16.88             & 17.74             & 17.92             & 17.78    & 20.71  &   17.58   \\
Lower Partial Standard Deviation (\%) & 17.56             & 17.83             & 17.63             & 17.49             & 17.58             & 18.94             & 19.16             & 19.13    & 21.49  &   18.48   \\
Kurtosis                              & 3.29              & 3.04              & 3.16              & 3.22              & 3.17              & 2.84              & 2.88              & 2.60     & 2.19   &   2.67    \\
Skewness                              & --0.25             & --0.30             & --0.32             & --0.31             & --0.31             & --0.32             & --0.35             & --0.33    & --0.34  &   --0.36   \\
Average Diversification Ratio         & 3.10              & 3.04              & 3.07              & 3.09              & 3.09              & 2.63              & 2.60              & 2.51     & 2.18   &   2.66    \\
Average Max. Weight                   & 0.18              & 0.19              & 0.19              & 0.19              & 0.19              & 0.18              & 0.18              & 0.18     & 0.13   &   0.19    \\
Average Min. Weight                   & 0.00              & 0.00              & 0.00              & 0.00              & 0.00              & 0.00              & 0.00              & 0.00     & 0.00   &   0.00    \\
Average Gross Leverage                & 1.00              & 1.00              & 1.00              & 1.00              & 1.00              & 1.00              & 1.00              & 1.00     & 1.00   &   1.00     \\
Proportion of Leverage (\%)           & 0.00              & 0.00              & 0.00              & 0.00              & 0.00              & 0.00              & 0.00              & 0.00     & 0.00   &   0.00     \\
Average Turnover (\%)                 & 0.17              & 0.16              & 0.16              & 0.16              & 0.16              & 0.03              & 0.03              & 0.04     & 0.03   &   0.07     \\
Average Excess Return (\%)            & 14.29             & 15.86             & 16.18             & 14.98             & 15.06             & 20.22             & 15.85             & 16.28    & 14.36  &   16.6     \\
Cumulative Return (\%)                & 27.49             & 31.30             & 32.15             & 29.25             & 29.44             & 42.18             & 30.91             & 32.04    & 26.01  &   32.92    \\
Sharpe Ratio                          & 0.84              & 0.93              & 0.95              & 0.89              & 0.89              & 1.14              & 0.88              & 0.92     & 0.69   &   0.94   \\
\hline
& \multicolumn{2}{c}{1 Factor}     & \multicolumn{2}{c}{3 Factors}  &   \multicolumn{2}{c}{5 Factors}    & \multicolumn{2}{c}{7 Factors}    \\
& \multicolumn{2}{c}{VHAR}         & \multicolumn{2}{c}{VHAR}     &     \multicolumn{2}{c}{VHAR}         & \multicolumn{2}{c}{VHAR}         \\
& \multicolumn{2}{c}{(Log matrix)} & \multicolumn{2}{c}{(Log matrix)} &    \multicolumn{2}{c}{(Log matrix)} & \multicolumn{2}{c}{(Log matrix)} \\
& LASSO             & adaLASSO          & LASSO             & adaLASSO    &      LASSO             & adaLASSO          & LASSO             & adaLASSO \\
\hline
Standard Deviation (\%)               & 16.96             & 16.98             & 16.55             & 16.59             & 16.34             & 16.47             & 16.31             & 16.44             \\
Lower Partial Standard Deviation (\%) & 17.51             & 17.64             & 17.29             & 17.27             & 16.88             & 17.10             & 16.89             & 17.03             \\
Kurtosis                              & 4.32              & 4.27              & 4.18              & 3.70             & 3.37              & 3.46              & 3.69              & 3.56              \\
Skewness                              & --0.35             & --0.35             & --0.36             & --0.27             & --0.19             & --0.21             & --0.20             & --0.23             \\
Average Diversification Ratio         & 3.10              & 3.09              & 3.09              & 3.10              & 3.12              & 3.13              & 3.11              & 3.11              \\
Average Max. Weight                   & 0.17              & 0.18              & 0.19              & 0.19              & 0.19              & 0.19              & 0.19              & 0.19              \\
Average Min. Weight                   & 0.00              & 0.00              & 0.00              & 0.00              & 0.00              & 0.00              & 0.00              & 0.00              \\
Average Gross Leverage                & 1.00              & 1.00              & 1.00              & 1.00              & 1.00              & 1.00              & 1.00              & 1.00              \\
Proportion of Leverage (\%)           & 0.00              & 0.00              & 0.00              & 0.00              & 0.00              & 0.00              & 0.00              & 0.00              \\
Average Turnover (\%)                 & 0.08              & 0.08              & 0.07              & 0.08              & 0.07              & 0.08              & 0.07              & 0.07              \\
Average Excess Return (\%)            & 17.60             & 17.57             & 17.62             & 18.04             & 18.02             & 18.17             & 17.13             & 17.04             \\
Cumulative Return (\%)                & 35.71             & 35.63             & 35.95             & 37.01             & 37.06             & 37.38             & 34.79             & 34.50             \\
Sharpe Ratio                          & 1.04              & 1.03              & 1.06              & 1.09              & 1.10              & 1.10              & 1.05              & 1.04              \\

\hline
\end{tabular}}
\end{threeparttable}
\end{table}
\end{landscape}

\section{Conclusion}\label{S:Conclusions}

We propose a model to forecast large realized covariance matrices of returns, applying it to some of the constituents of the S\&P 500 on a daily basis. To address the curse of dimensionality, we decompose the return covariance matrix using standard firm-level factors (e.g., size, value, and profitability) and use sectoral restrictions in the residual covariance matrix. This restricted model is then estimated using vector heterogeneous autoregressive (VHAR) models with the least absolute shrinkage and selection operator (LASSO). Our methodology improves forecasting precision relative to standard benchmarks and leads to better estimates of minimum variance portfolios.


\newpage

\section*{Appendix}\appendix

\setcounter{table}{0}
\renewcommand{\thetable}{A\arabic{table}}

\setcounter{figure}{0}
\renewcommand{\thefigure}{A\arabic{figure}}

\setcounter{equation}{0}
\renewcommand{\theequation}{A\arabic{equation}}

\section{Factor Loadings}\label{appendix_factor_loadings}

This appendix describes how daily loadings on factors can be calculated using daily realized covariance matrices of returns $\bs\Sigma_{t}$, daily factor covariance matrices $\bs\Sigma_{f,t}$, and factor weights $\bs W$ (which are calculated yearly). To simplify the notation, we drop the subscript $t$. We derive the equation for a daily matrix of loadings $\widehat{\bs B}$ (the final equation is applied to daily data, providing 1,495 matrices $\widehat{\bs B}$). Letters with one subscript $t$ represent vectors with cross-sectional data at time $t$, while letters with one superscript $n$ represent time series data for asset $n$.

Assume that we have $T$ observations for $K$ factors and $N$ assets (in our case, $T$ can be considered intra-daily high-frequency observations). Stacking all observations in matrix form,
\begin{equation}
\begin{gathered}
\bs F =
\begin{pmatrix}
f_{1,1} & \hdots & f_{K,1} \\
\vdots     & \ddots    & \vdots \\
f_{1,T} & \hdots & f_{K,T}
\end{pmatrix}
=
\begin{pmatrix}
\bs f_{1}' \\
\vdots \\
\bs f_{T}'
\end{pmatrix}
\\
\bs R =
\begin{pmatrix}
r_{1,1} & \hdots & r_{1,T} \\
\vdots     & \ddots    & \vdots \\
r_{N,1} & \hdots & r_{N,T}
\end{pmatrix}
=
\begin{pmatrix}
\bs r_{1}' \\
\vdots \\
\bs r_{N}'
\end{pmatrix}
 =
\begin{pmatrix}
\bs r_{1} & \hdots & \bs r_{T}
\end{pmatrix}.
\end{gathered}
\end{equation}

From equation (\ref{factors_components}), we can rewrite the matrix $\bs F$ as
\begin{equation}\label{factors_matrix}
\begin{gathered}
\bs F =
\begin{pmatrix}
\bs w_{1}'\bs r_{1} & \hdots & \bs w_{K}'\bs r_{1} \\
\vdots     & \ddots    & \vdots \\
\bs w_{1}'\bs r_{T} & \hdots & \bs w_{K}'\bs r_{T}
\end{pmatrix}
=
\begin{pmatrix}
\bs r_{1}'\bs w_{1} & \hdots & \bs r_{1}'\bs w_{K} \\
\vdots     & \ddots    & \vdots \\
\bs r_{T}'\bs w_{1} & \hdots & \bs r_{T}'\bs w_{K}
\end{pmatrix}
=
\begin{pmatrix}
\bs r_{1}' \\ \vdots \\ \bs r_{T}'
\end{pmatrix}
\begin{pmatrix}
\bs w_{1} & \hdots & \bs w_{K}
\end{pmatrix}
= \bs R'\bs W.
\end{gathered}
\end{equation}

With this setup, consider the linear model of asset $n$, $n\in\{1,\ldots,N\}$, on $K$ factors:
\begin{equation}
\begin{gathered}
\underset{\bs r^{n}}{\begin{pmatrix}
r_{n,1} \\
\vdots \\
r_{n,T}
\end{pmatrix}}
=
\underset{\bs F}{\begin{pmatrix}
\bs f_{1}' \\
\vdots \\
\bs f_{T}'
\end{pmatrix}}
\underset{\bs b_{n}}{
\begin{pmatrix}
b_{n,1} \\
\vdots \\
b_{n,K}
\end{pmatrix}}
+
\begin{pmatrix}
\epsilon_{n,1} \\
\vdots \\
\epsilon_{n,T}
\end{pmatrix}.
\end{gathered}
\end{equation}

The OLS estimator for $\bs b_{n}$, $\widehat{\bs b}_{n}$, is given by
\begin{equation}
\widehat{\bs b}_{n}=(\bs F'\bs F)^{-1}\bs F'\bs r^{n} \quad \quad \textnormal{for}\quad i = 1,\ldots,N.
\end{equation}

In matrix form for all $n$,
\begin{equation}
\begin{gathered}
\widehat{\bs B}=
\begin{pmatrix}
\widehat{\bs b}_{1} & \hdots & \widehat{\bs b}_{N}
\end{pmatrix}
=
(\bs F'\bs F)^{-1}\bs F'
\begin{pmatrix}
\bs r^{1} & \hdots & \bs r^{N}
\end{pmatrix} = \\
=(\bs F'\bs F)^{-1}\bs F'
\begin{pmatrix}
\bs r_{1}' \\
\vdots \\
\bs r_{T}'
\end{pmatrix}
=(\bs F'\bs F)^{-1}\bs F'\bs R' =\\
=(\bs F'\bs F)^{-1}\bs W'\bs R\bs R'
\xrightarrow{mean = 0} (\Sigma_{f})^{-1}W'\Sigma,
\end{gathered}
\end{equation}
with equation (\ref{factors_matrix}) being used on the last line. Again, in our setup, $\widehat{\bs B}$ is different for every day in the sample.

\newpage

\section{Betas' Long Memory}\label{appendix_betas_long_memory}
\begin{figure}[H]
\caption{\textbf{Estimated Fractional Differencing Parameter ($d$) - Whittle Method}}
\begin{footnotesize}
\begin{spacing}{}
These panels show the distribution of fractional differencing parameter ($d$) estimated by the Whittle method. Each panel corresponds to one of the factor configurations (1, 3, 5, or 7 factors). There are $430 \times \textnormal{number of factors}$ series of betas for each case. We plot the distribution of d for these series.
\bigskip
\end{spacing}
\end{footnotesize}
\includegraphics[width =1\textwidth]{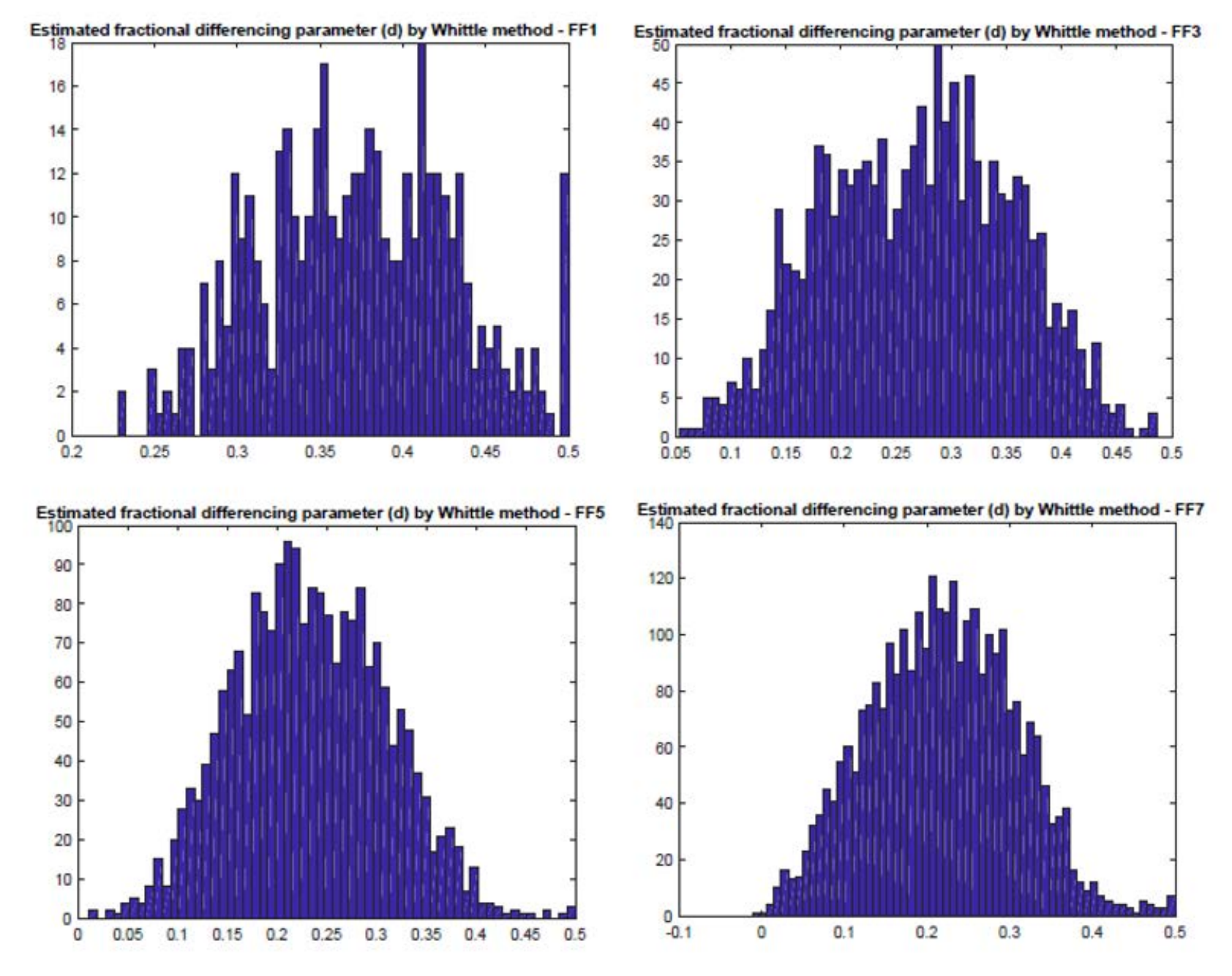}
\label{fig_whittle}
\end{figure}

This section investigates the long-range dependency for the series of factor loadings (``betas''). For this, we estimate the fractional differencing parameter ($d$) distribution for four groups of betas, corresponding to the 1F, 3F, 5F, and 7F factor configurations. For each factor configuration, we pool all series into a single distribution (e.g., for FF3, we plot the distribution of $d$ for the 430 x 3 series). Figure \ref{fig_whittle} shows the results obtained using the Whittle method. From the results, we see that, in all four cases, a few series have values of $d$ close to zero, which evidences long memory.

\begin{figure}[H]
\caption{\textbf{Estimated Fractional Differencing Parameter ($d$) - GPH Method}}
\begin{footnotesize}
\begin{spacing}{}
These panels show the distribution of fractional differencing parameter ($d$) estimated by the GPH method. Each panel corresponds to one of the factor configurations (1, 3, 5, or 7 factors). There are $430 \times \textnormal{number of factors}$ series of betas for each case. We plot the distribution of d for these series.
\bigskip
\end{spacing}
\end{footnotesize}
\includegraphics[width =1\textwidth]{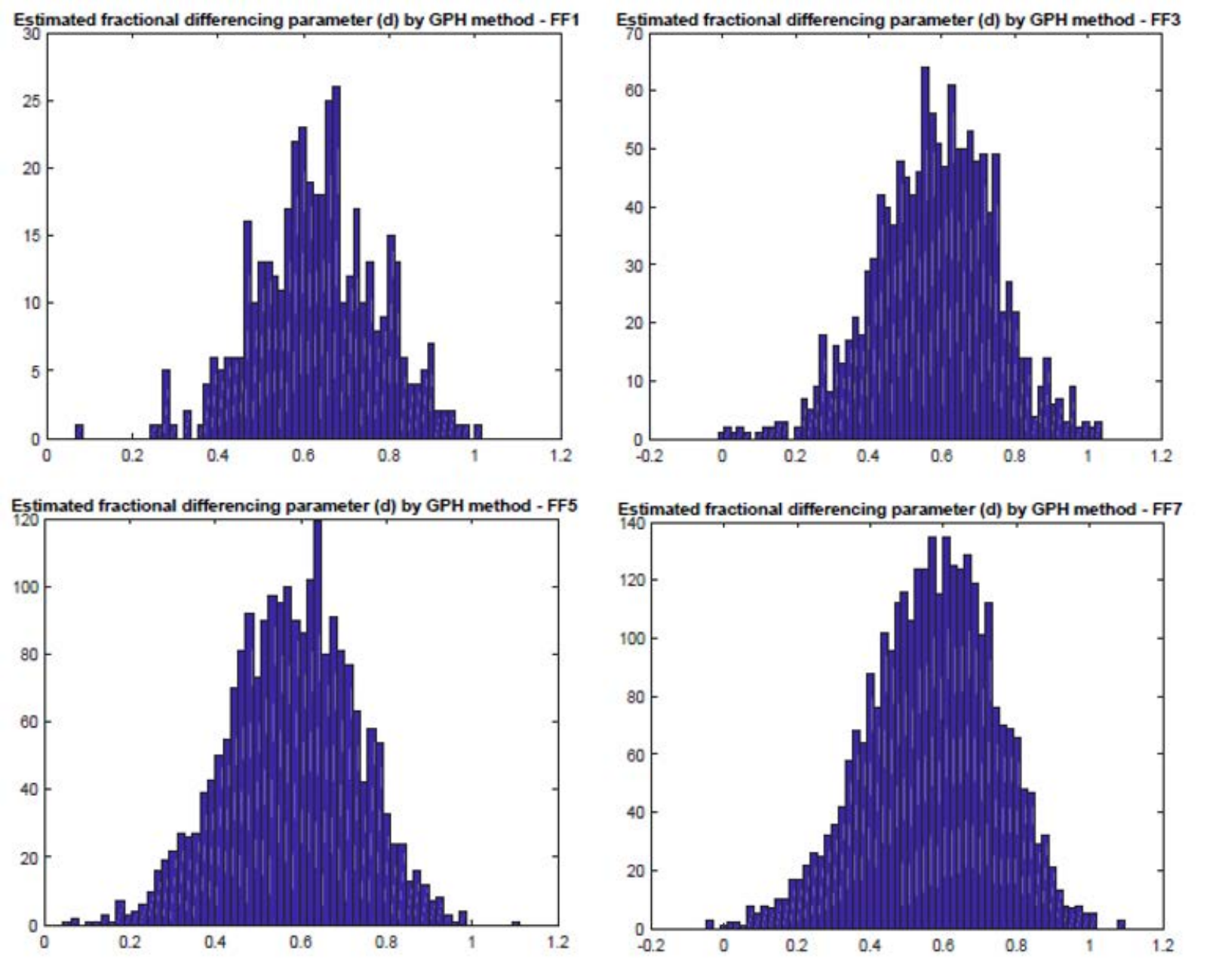}
\label{fig_gph}
\end{figure}

Figure \ref{fig_gph} shows similar results for the GPH (Geweke and Porter-Hudak) estimator. Again, in all cases, most series have a value of $d$ different from zero. We can also see that the distributions are centered close to 0.5 and away from 1, which indicates stationarity.

\newpage

\section{Factor Construction}\label{appendix_factors_construction}

\textbf{Market}: The market portfolio is a value-weighted portfolio of all stocks in our sample. The market factor is the excess return of this portfolio relative to the risk-free rate.

\noindent\textbf{Size and Value}: These factors are based on a double sorting of stocks on market equity and book-to-market equity. Market equity is defined as the last available price in June and the corresponding value of outstanding shares. The book-to-market ratio is calculated using the book value of the previous fiscal year and the market equity from the last day of December of the previous year. Stocks are ranked on market equity and split into small and big portfolios (S and B). Book-to-market is used to split the stocks into three book-to-market equity groups based on breakpoints for the bottom 30\% (low, or L), middle 40\% (medium, or M), and high 30\% (high, or H). Six portfolios are formed from intersections of the groups mentioned above: S/L, S/M, S/H, B/L, B/M, and B/H. Daily value-weighted returns are calculated for the six portfolios for the whole year (portfolios are rebalanced annually). The factor SMB (small minus big) is the difference between the simple average of returns on the three small-stock portfolios and the simple average of returns on the three big-stock portfolios, that is,
\[
SMB = \frac{(R_{S/L} + R_{S/M} + R_{S/H})}{3} - \frac{(R_{B/L} + R_{B/M} + R_{B/H})}{3}.
\]

Similarly, the factor HML (high minus low) is defined as the daily difference of the simple average of returns on the two high book-to-market portfolios (S/H and B/H) and the simple average of returns on the two low book-to-market portfolios (S/L and B/L), that is,

\[
HML = \frac{(R_{S/H} + R_{B/H})}{2} - \frac{(R_{S/L} + R_{B/L})}{2}.
\]

\noindent\textbf{Gross Profitability}: The factor construction follows \citet{novy2013other}. The signal gross profitability (GP) is the ratio between gross profits and total assets. Stocks are ranked and split in 10 deciles every year using current-year financial data. The gross profitability portfolio consists of a long strategy in the group with the lowest GP and a short in the group with the highest GP. For both groups, value-weighted returns are calculated, and the factor GP is given by
\[
GP = R_{lowGP} - R_{highGP}.
\]

\noindent\textbf{Accruals}: This factor follows \citet{sloan1996}. The signal accrual is defined as
\[
\textnormal{Accruals} = \frac{\Delta ACT - \Delta CHE - \Delta LCT + \Delta DLC + \Delta TXP -DP}{(AT + AT_{-12})/2},
\]
where $ACT$ is the annual total current assets, $CHE$ is the annual total cash and short-term investments, $LCT$ is the annual current liabilities, $DLC$ is the annual debt in current liabilities, $TXP$ is the annual income taxes payable, $DP$ is the annual depreciation and amortization, and $(AT + AT_{-12})/2$ represents the average total assets over the last two years. $\Delta$ stands for the annual variation in these variables. Stocks are split into 10 deciles yearly, and two value-weighted portfolios are formed on the lowest and highest deciles. The accruals portfolio consists of a strategy that is long on the small accrual portfolio and short on the high accrual portfolio. The factor is then given by
\[
Accrual = R_{lowAccrual} - R_{high_accrual}.
\]

\noindent\textbf{Asset Growth}: This factor follows \citet{cooper2008asset}. The asset growth signal is defined as $\textnormal{Asset Growth} = AT/AT_{-12}$, with $AT$ and $AT_{-12}$ previously defined in the accruals section. Analogous to the procedure used for the accruals factor, the asset growth portfolio is simply a strategy that is long on the lowest decile asset growth portfolio and short on the highest decile portfolio (the stocks are sorted into 10 deciles). Following this procedure, the factor is then given by
\[
Asset Growth = R_{lowAssetGrowth} - R_{highAssetGrowth}.
\]

\noindent\textbf{Investment}: The factor construction follows \citet{lyandres2008new}. The investment signal is defined as
\[
\textnormal{Investment} = (\Delta PPEGT + \Delta INVT)/AT_{-12},
\]
where $PPEGT$ is the gross total property, plant, and equipment (COMPUSTAT's variable 'ppegt'), and $INVT$ is total inventories (COMPUSTAT's 'invt'). For the investment portfolio, the stocks are triple-sorted in size, value (as in the Fama-French factors), and investment. For each of the three characteristics, each stock is classified into one of three groups: low (30\%), medium (30\%-70\%), or high (70\%-100\%). This procedure results in 27 different portfolios. The investment portfolio consists of a strategy that is long on the low investment portfolio (the simple average between the nine groups with low investment) and short on the high investment portfolio (the simple average between the nine groups with high investment), that is,
\[
Investment = R_{lowInvestment} - R_{highInvestment}.
\]

\noindent\textbf{Risk-Free}: We obtain the daily series of returns from Kenneth French's website.

\newpage

\section{Diagnostic Criterion for Approximate Factor Structure}\label{appendix_approximate_factor}
In this appendix, we follow the procedure described in \citet{gagliardini2018diagnostic} to check for the existence of omitted factors in our models. The diagnostic criterion for omitted factors is based on the following expression:
\[
\xi(k) = \mu_{k+1}\left(\frac{1}{NT}\sum_{i}\bs\epsilon_{i}\bs\epsilon_{i}'\right) - g(N,T),
\]
where \(\mu_{k+1}(.)\) denotes the $(k+1)$-th largest eigenvalue of a symmetric matrix, $N$ represents the number of assets, $T$ represents the sample size, and $\bs\epsilon_{i}$ is a vector of dimension $T\times 1$ with the residuals from the $i$-th asset relative to the model being tested. $g(N,T)$ is a penalty function given by $g(N,T)=\left[\frac{N+T}{NT}\log(\frac{NT}{N+T})\right]\widehat{\sigma}^{2}$, where $\widehat{\sigma}^2$ is the estimated variance of the residuals. The criterion detects $k$ omitted factors in the residuals when $\xi(k) < 0$. To illustrate, for a given specification, if $\xi(2)>0$ and $\xi(3)<0$, the residuals present a factor structure with two omitted factors.

\begin{figure}[H]
    \caption{Approximate Factor Structure for the 1F, 3F, 5F, and 7F Models}
        \begin{footnotesize}
            \begin{spacing}{}
                \bigskip
            \end{spacing}
         \end{footnotesize}
    \centering
    \includegraphics[width = 0.9\textwidth]{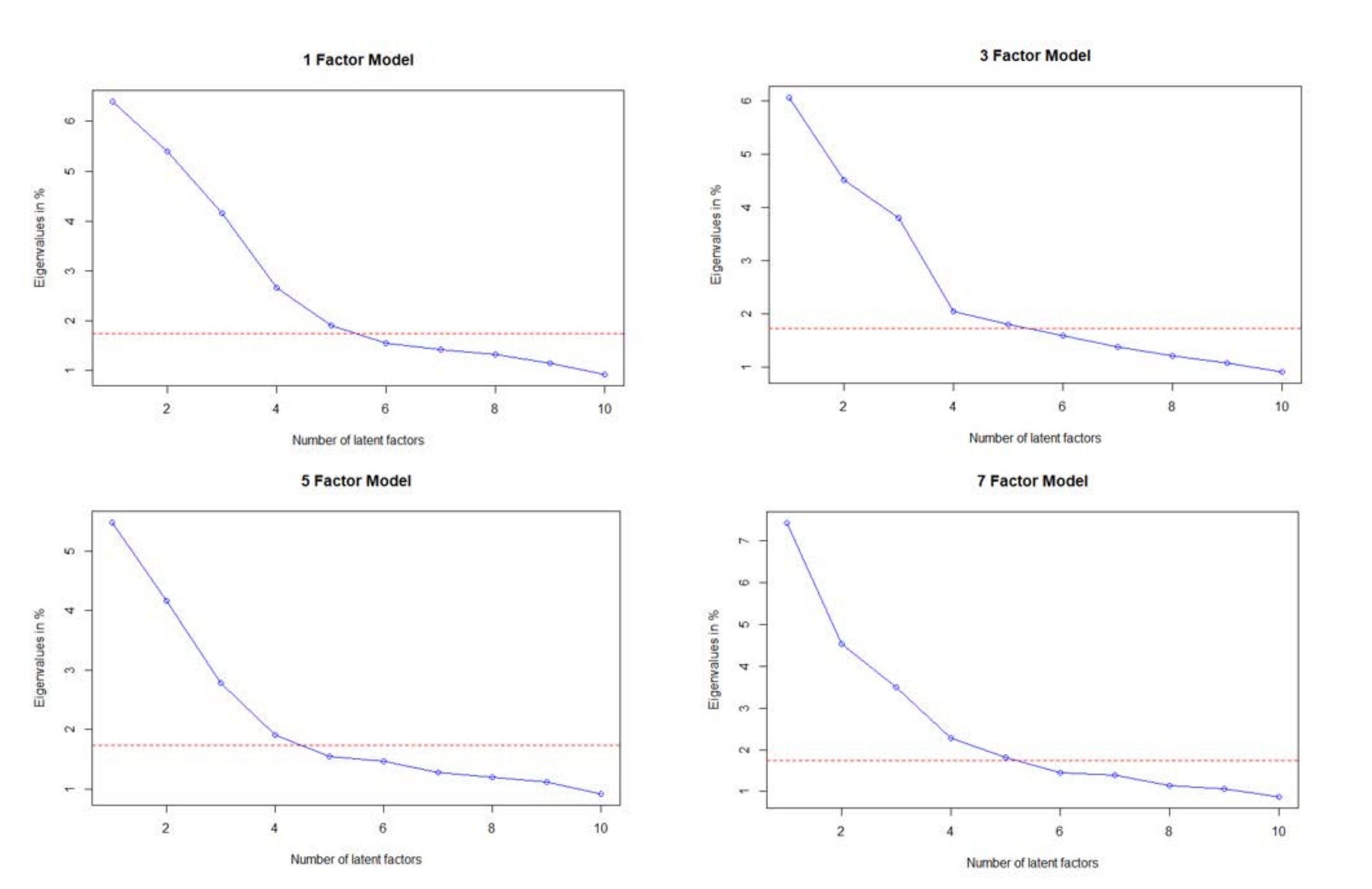}
    \label{fig_scaillet_1}
\end{figure}

\begin{figure}[H]
    \caption{Approximate Factor Structure for the 7F + 10 Industry Factors Model}
        \begin{footnotesize}
            \begin{spacing}{}
                \bigskip
            \end{spacing}
         \end{footnotesize}
    \centering
    \includegraphics[width = 0.7\textwidth]{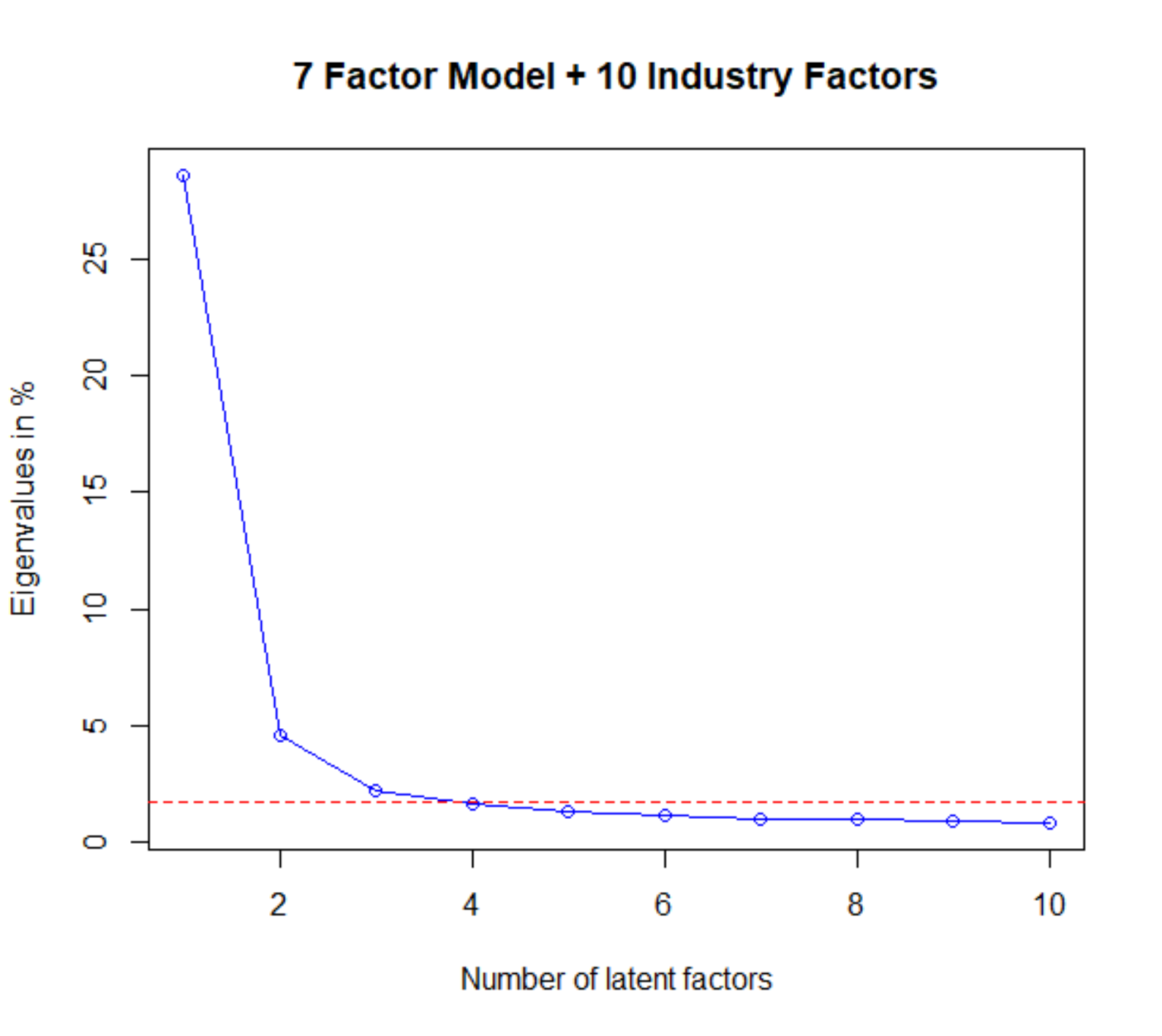}
    \label{fig_scaillet_2}
\end{figure}

In Figure \ref{fig_scaillet_1}, we show the values of $\xi(k)$ for four factor configurations: 1F, 3F, 5F, and 7F. In all cases, we detect the presence of omitted factors, which suggests that imposing more structure on the residuals is appropriate. Since we used sector classifications to restrict the residual dynamics, we also evaluated the criterion in one of these cases. In Figure \ref{fig_scaillet_2}, we show similar results for the better performing model of Section \ref{sec_portfolio}, that is, the 7F with block diagonal residual covariance matrices. From the results, we verify a reduction in omitted factors from five to three in this case. 

\section{Residual Covariance Matrix - Parameter Selection}\label{appendix_residual_parameter_selection}
In this appendix, we show the average equation size and average parameter change for models that forecast the residual covariance matrices. Since the number of categories of equations is much higher than the one used for the factor covariance matrices, we present the results in percentiles of the distribution (instead of individual plots).

\begin{table}[H]
\caption{\textbf{Average Equation Size Distribution for Blocks in the Residual Covariance Matrix (1-Factor VHAR).}}
\label{table_average_size_block_1f}
\centering
\begin{minipage}{\linewidth}
\begin{footnotesize}
The distribution represents the percentiles and means values calculated over 473 estimated models. The 430 stocks are separated into ten groups, according to SIC. Averages are calculated among stocks in the same group.
\end{footnotesize}
\end{minipage}
\resizebox{\columnwidth}{!}{
\begin{threeparttable}
\bigskip
\begin{tabular}{lcccccc}  \hline \Tstrut
                  Distribution                        & \multicolumn{1}{l}{Min} & \multicolumn{1}{l}{0.25} & \multicolumn{1}{l}{0.5} & \multicolumn{1}{l}{0.75} & \multicolumn{1}{l}{Max} & \multicolumn{1}{l}{Mean} \Bstrut \\  \hline
                                          & \multicolumn{1}{l}{}    & \multicolumn{1}{l}{}     & \multicolumn{1}{l}{}    & \multicolumn{1}{l}{}     & \multicolumn{1}{l}{}    & \multicolumn{1}{l}{}  \\
                                          & \multicolumn{6}{c}{Variance Equations}                                                                                                                       \\  \cline{2-7}
\textit{Group}                            & \multicolumn{1}{l}{}    & \multicolumn{1}{l}{}     & \multicolumn{1}{l}{}    & \multicolumn{1}{l}{}     & \multicolumn{1}{l}{}    & \multicolumn{1}{l}{}     \\
Consumer Non-Durables                     & 11.29 & 11.90 & 12.16 & 12.48 & 13.55 & 12.25 \\
Consumer Durables                         & 6.12  & 6.38  & 6.62  & 6.88  & 7.62  & 6.69  \\
Manufacturing                             & 16.05 & 16.57 & 16.74 & 16.92 & 17.26 & 16.74 \\
Oil, Gas, and Coal Extraction             & 9.38  & 10.16 & 10.47 & 10.66 & 11.38 & 10.43 \\
Business Equipment                        & 14.70 & 15.15 & 15.33 & 15.52 & 15.97 & 15.33 \\
Telecommunications                        & 6.50  & 6.90  & 7.40   & 7.60   & 7.90   & 7.28  \\
Wholesale and Retail                      & 12.51 & 13.27 & 13.44 & 13.6  & 14.00    & 13.43 \\
Health Care, Medical Equipments, and Drugs & 11.69 & 12.19 & 12.46 & 12.85 & 13.35 & 12.5  \\
Utilities                                 & 11.86 & 12.56 & 12.83 & 13.17 & 13.72 & 12.85 \\
Others                                    & 15.46 & 15.88 & 16.52 & 16.97 & 18.1  & 16.47                    \\
                                          & \multicolumn{1}{l}{}    & \multicolumn{1}{l}{}     & \multicolumn{1}{l}{}    & \multicolumn{1}{l}{}     & \multicolumn{1}{l}{}    & \multicolumn{1}{l}{}     \\
                                          & \multicolumn{6}{c}{Covariance Equations}                                                                                                                     \\   \cline{2-7}
\textit{Group}                            & \multicolumn{1}{l}{}    & \multicolumn{1}{l}{}     & \multicolumn{1}{l}{}    & \multicolumn{1}{l}{}     & \multicolumn{1}{l}{}    & \multicolumn{1}{l}{}     \\
Consumer Non-Durables                     & 0.77 & 1.07 & 1.51 & 1.63 & 1.75 & 1.36 \\
Consumer Durables                         & 0.39 & 0.96 & 1.50 & 1.68 & 1.93 & 1.35 \\
Manufacturing                             & 0.51 & 0.60 & 0.69 & 0.76 & 0.82 & 0.68 \\
Oil, Gas, and Coal Extraction             & 1.72 & 2.03 & 2.43 & 2.59 & 3.00 & 2.33 \\
Business Equipment                        & 0.37 & 0.45 & 0.60 & 0.64 & 0.69 & 0.55 \\
Telecommunications                        & 1.44 & 1.67 & 1.78 & 1.89 & 2.18 & 1.79 \\
Wholesale and Retail                      & 0.56 & 0.69 & 1.03 & 1.14 & 1.28 & 0.95 \\
Health Care, Medical Equipments, and Drugs & 0.58 & 0.85 & 1.39 & 1.83 & 2.02 & 1.34 \\
Utilities                                 & 0.69 & 0.77 & 0.84 & 1.48 & 1.92 & 1.09 \\
Others                                    & 0.48 & 0.54 & 0.64 & 0.74 & 0.77 & 0.63    \Bstrut   \\   \hline
\end{tabular}
\end{threeparttable}}
\end{table}

\begin{table}[H]
\caption{\textbf{Average Parameter Change Distribution for Blocks in the Residual Covariance Matrix (1-Factor VHAR).}}
\label{table_average_change_block_1f}
\centering
\begin{minipage}{\linewidth}
\begin{footnotesize}
The distribution represents the percentiles and means values calculated over 473 estimated models. The 430 stocks are separated into ten groups, according to SIC. Averages are calculated among stocks in the same group.
\end{footnotesize}
\end{minipage}
\resizebox{\columnwidth}{!}{
    \begin{threeparttable}
\bigskip
\begin{tabular}{lcccccc}  \hline \Tstrut
                  Distribution                        & \multicolumn{1}{l}{Min} & \multicolumn{1}{l}{0.25} & \multicolumn{1}{l}{0.5} & \multicolumn{1}{l}{0.75} & \multicolumn{1}{l}{Max} & \multicolumn{1}{l}{Mean} \Bstrut \\  \hline
                                          & \multicolumn{1}{l}{}    & \multicolumn{1}{l}{}     & \multicolumn{1}{l}{}    & \multicolumn{1}{l}{}     & \multicolumn{1}{l}{}    & \multicolumn{1}{l}{}  \\
                                          & \multicolumn{6}{c}{Variance Equations}                                                                                                                       \\  \cline{2-7}
\textit{Group}                            & \multicolumn{1}{l}{}    & \multicolumn{1}{l}{}     & \multicolumn{1}{l}{}    & \multicolumn{1}{l}{}     & \multicolumn{1}{l}{}    & \multicolumn{1}{l}{}     \\
Consumer Non-Durables                     & 0.21 & 0.94 & 1.25 & 1.66 & 3.75 & 1.33 \\
Consumer Durables                         & 0.00 & 0.00 & 0.00 & 1.56 & 4.69 & 0.52 \\
Manufacturing                             & 0.17 & 0.78 & 0.99 & 1.25 & 3.48 & 1.05 \\
Oil, Gas, and Coal Extraction             & 0.10 & 0.88 & 1.37 & 1.86 & 4.30 & 1.43 \\
Business Equipment                        & 0.46 & 0.96 & 1.16 & 1.43 & 4.03 & 1.21 \\
Telecommunications                        & 0.00 & 0.00 & 0.00 & 1.00 & 4.00 & 0.65 \\
Wholesale and Retail                      & 0.20 & 0.89 & 1.14 & 1.43 & 2.91 & 1.19 \\
Health Care, Medical Equipment, and Drugs & 0.00 & 0.74 & 1.04 & 1.48 & 6.07 & 1.16 \\
Utilities                                 & 0.15 & 0.62 & 0.93 & 1.16 & 4.32 & 0.94 \\
Others                                    & 0.37 & 0.68 & 0.79 & 0.94 & 1.81 & 0.82 \\
                                          & \multicolumn{1}{l}{}    & \multicolumn{1}{l}{}     & \multicolumn{1}{l}{}    & \multicolumn{1}{l}{}     & \multicolumn{1}{l}{}    & \multicolumn{1}{l}{}     \\
                                          & \multicolumn{6}{c}{Covariance Equations}                                                                                                                     \\   \cline{2-7}
\textit{Group}                            & \multicolumn{1}{l}{}    & \multicolumn{1}{l}{}     & \multicolumn{1}{l}{}    & \multicolumn{1}{l}{}     & \multicolumn{1}{l}{}    & \multicolumn{1}{l}{}     \\
Consumer Non-Durables                     & 0.10 & 0.30 & 0.39 & 0.48 & 0.96 & 0.39 \\
Consumer Durables                         & 0.00 & 0.45 & 0.89 & 1.34 & 4.46 & 0.94 \\
Manufacturing                             & 0.07 & 0.11 & 0.13 & 0.14 & 0.24 & 0.13 \\
Oil, Gas, and Coal Extraction             & 0.26 & 0.55 & 0.67 & 0.79 & 1.32 & 0.68 \\
Business Equipment                        & 0.05 & 0.09 & 0.10 & 0.12 & 0.17 & 0.1  \\
Telecommunications                        & 0.00 & 0.44 & 0.89 & 1.33 & 3.56 & 0.9  \\
Wholesale and Retail                      & 0.07 & 0.17 & 0.21 & 0.26 & 0.42 & 0.21 \\
Health Care, Medical Equipment, and Drugs & 0.11 & 0.30 & 0.43 & 0.58 & 1.04 & 0.45 \\
Utilities                                 & 0.06 & 0.22 & 0.31 & 0.41 & 0.78 & 0.32 \\
Others                                    & 0.04 & 0.06 & 0.07 & 0.07 & 0.1  & 0.07    \Bstrut   \\   \hline
\end{tabular}
\end{threeparttable}}
\end{table}

\begin{table}[H]
\caption{\textbf{Average Equation Size Distribution for Blocks in the Residual Covariance Matrix (3-Factor VHAR).}}
\label{table_average_size_block_3f}
\centering
\begin{minipage}{\linewidth}
\begin{footnotesize}
The distribution represents the percentiles and means values calculated over 473 estimated models. The 430 stocks are separated into ten groups, according to SIC. Averages are calculated among stocks in the same group.
\end{footnotesize}
\end{minipage}
\resizebox{\columnwidth}{!}{
\begin{threeparttable}
\bigskip
\begin{tabular}{lcccccc}  \hline \Tstrut
                  Distribution                        & \multicolumn{1}{l}{Min} & \multicolumn{1}{l}{0,25} & \multicolumn{1}{l}{0,5} & \multicolumn{1}{l}{0,75} & \multicolumn{1}{l}{Max} & \multicolumn{1}{l}{Mean} \Bstrut \\  \hline
                                          & \multicolumn{1}{l}{}    & \multicolumn{1}{l}{}     & \multicolumn{1}{l}{}    & \multicolumn{1}{l}{}     & \multicolumn{1}{l}{}    & \multicolumn{1}{l}{}  \\
                                          & \multicolumn{6}{c}{Variance Equations}                                                                                                                       \\  \cline{2-7}
\textit{Group}                            & \multicolumn{1}{l}{}    & \multicolumn{1}{l}{}     & \multicolumn{1}{l}{}    & \multicolumn{1}{l}{}     & \multicolumn{1}{l}{}    & \multicolumn{1}{l}{}     \\
Consumer Non-Durables                     & 11.19                   & 11.97                    & 12.19                   & 12.48                    & 13.61                   & 12.29                    \\
Consumer Durables                         & 6.25                    & 6.50                     & 6.62                    & 7.00                     & 7.75                    & 6.79                     \\
Manufacturing                             & 16.22                   & 16.69                    & 16.85                   & 17.05                    & 17.46                   & 16.86                    \\
Oil, Gas, and Coal Extraction             & 9.00                    & 9.78                     & 10.19                   & 10.59                    & 11.12                   & 10.17                    \\
Business Equipment                        & 15.03                   & 15.46                    & 15.64                   & 15.77                    & 16.18                   & 15.62                    \\
Telecommunications                        & 6.60                    & 6.90                     & 7.30                    & 7.60                     & 7.90                    & 7.27                     \\
Wholesale and Retail                      & 12.80                   & 13.36                    & 13.53                   & 13.71                    & 14.22                   & 13.55                    \\
Health Care, Medical Equipment, and Drugs & 11.54                   & 12.19                    & 12.42                   & 12.92                    & 13.58                   & 12.53                    \\
Utilities                                 & 12.03                   & 12.75                    & 13.00                   & 13.25                    & 14.25                   & 13.03                    \\
Others                                    & 15.41                   & 15.85                    & 16.47                   & 17.13                    & 18.11                   & 16.53                    \\
                                          & \multicolumn{1}{l}{}    & \multicolumn{1}{l}{}     & \multicolumn{1}{l}{}    & \multicolumn{1}{l}{}     & \multicolumn{1}{l}{}    & \multicolumn{1}{l}{}     \\
                                          & \multicolumn{6}{c}{Covariance Equations}                                                                                                                     \\   \cline{2-7}
\textit{Group}                            & \multicolumn{1}{l}{}    & \multicolumn{1}{l}{}     & \multicolumn{1}{l}{}    & \multicolumn{1}{l}{}     & \multicolumn{1}{l}{}    & \multicolumn{1}{l}{}     \\
Consumer Non-Durables                     & 0.66                    & 0.83                     & 1.28                    & 1.39                     & 1.52                    & 1.15                     \\
Consumer Durables                         & 0.21                    & 0.36                     & 0.54                    & 0.68                     & 0.89                    & 0.53                     \\
Manufacturing                             & 0.47                    & 0.55                     & 0.63                    & 0.68                     & 0.72                    & 0.61                     \\
Oil, Gas, and Coal Extraction             & 1.46                    & 1.76                     & 2.09                    & 2.23                     & 2.64                    & 2.01                     \\
Business Equipment                        & 0.34                    & 0.40                     & 0.55                    & 0.67                     & 0.71                    & 0.54                     \\
Telecommunications                        & 1.36                    & 1.53                     & 1.62                    & 1.73                     & 2.11                    & 1.65                     \\
Wholesale and Retail                      & 0.51                    & 0.60                     & 0.89                    & 1.02                     & 1.09                    & 0.83                     \\
Health Care, Medical Equipment, and Drugs & 0.42                    & 0.73                     & 1.12                    & 1.58                     & 1.77                    & 1.13                     \\
Utilities                                 & 0.67                    & 0.74                     & 0.85                    & 1.58                     & 2.06                    & 1.14                     \\
Others                                    & 0.52                    & 0.60                     & 0.65                    & 0.70                     & 0.72                    & 0.64    \Bstrut   \\   \hline
\end{tabular}
\end{threeparttable}}
\end{table}

\begin{table}[H]
\caption{\textbf{Average Parameter Change Distribution for Blocks in the Residual Covariance Matrix (3-Factor VHAR).}}
 \label{table_average_change_block_3f}
\centering
\begin{minipage}{\linewidth}
\begin{footnotesize}
The distribution represents the percentiles and means values calculated over 473 estimated models. The 430 stocks are separated into ten groups, according to SIC. Averages are calculated among stocks in the same group.
\end{footnotesize}
\end{minipage}
\resizebox{\columnwidth}{!}{
\begin{threeparttable}
\bigskip
\begin{tabular}{lcccccc}  \hline \Tstrut
                  Distribution                        & \multicolumn{1}{l}{Min} & \multicolumn{1}{l}{0,25} & \multicolumn{1}{l}{0,5} & \multicolumn{1}{l}{0,75} & \multicolumn{1}{l}{Max} & \multicolumn{1}{l}{Mean} \Bstrut \\  \hline
                                          & \multicolumn{1}{l}{}    & \multicolumn{1}{l}{}     & \multicolumn{1}{l}{}    & \multicolumn{1}{l}{}     & \multicolumn{1}{l}{}    & \multicolumn{1}{l}{}  \\
                                          & \multicolumn{6}{c}{Variance Equations}                                                                                                                       \\  \cline{2-7}
\textit{Group}                            & \multicolumn{1}{l}{}    & \multicolumn{1}{l}{}     & \multicolumn{1}{l}{}    & \multicolumn{1}{l}{}     & \multicolumn{1}{l}{}    & \multicolumn{1}{l}{}     \\
Consumer Non-Durables                     & 0.21                 & 0.94 & 1.25 & 1.66 & 3.85 & 1.33 \\
Consumer Durables                         & 0.00                 & 0.00 & 0.00 & 1.56 & 3.12 & 0.52 \\
Manufacturing                             & 0.33                 & 0.83 & 1.02 & 1.23 & 3.36 & 1.08 \\
Oil, Gas, and Coal Extraction             & 0.00                 & 0.68 & 1.17 & 1.56 & 4.49 & 1.19 \\
Business Equipment                        & 0.40                 & 0.94 & 1.18 & 1.40 & 3.60 & 1.21 \\
Telecommunications                        & 0.00                 & 0.00 & 0.00 & 1.00 & 5.00 & 0.57 \\
Wholesale and Retail                      & 0.25                 & 0.94 & 1.19 & 1.43 & 3.80 & 1.24 \\
Health Care, Medical Equipment, and Drugs & 0.00                 & 0.74 & 1.04 & 1.48 & 4.88 & 1.16 \\
Utilities                                 & 0.08                 & 0.62 & 0.93 & 1.23 & 4.55 & 0.98 \\
Others                                    & 0.31                 & 0.67 & 0.80 & 0.94 & 1.63 & 0.82 \\
                                          & \multicolumn{1}{l}{}    & \multicolumn{1}{l}{}     & \multicolumn{1}{l}{}    & \multicolumn{1}{l}{}     & \multicolumn{1}{l}{}    & \multicolumn{1}{l}{}     \\
                                          & \multicolumn{6}{c}{Covariance Equations}                                                                                                                     \\   \cline{2-7}
\textit{Group}                            & \multicolumn{1}{l}{}    & \multicolumn{1}{l}{}     & \multicolumn{1}{l}{}    & \multicolumn{1}{l}{}     & \multicolumn{1}{l}{}    & \multicolumn{1}{l}{}     \\
Consumer Non-Durables                     & 0.10                 & 0.25 & 0.33 & 0.40 & 0.64 & 0.33 \\
Consumer Durables                         & 0.00                 & 0.00 & 0.45 & 0.89 & 3.12 & 0.50 \\
Manufacturing                             & 0.06                 & 0.09 & 0.11 & 0.12 & 0.20 & 0.11 \\
Oil, Gas, and Coal Extraction             & 0.15                 & 0.45 & 0.56 & 0.69 & 1.16 & 0.58 \\
Business Equipment                        & 0.04                 & 0.08 & 0.10 & 0.12 & 0.20 & 0.10 \\
Telecommunications                        & 0.00                 & 0.44 & 0.67 & 1.11 & 3.33 & 0.78 \\
Wholesale and Retail                      & 0.07                 & 0.15 & 0.18 & 0.22 & 0.42 & 0.19 \\
Health Care, Medical Equipment, and Drugs & 0.05                 & 0.25 & 0.34 & 0.45 & 0.86 & 0.36 \\
Utilities                                 & 0.08                 & 0.21 & 0.31 & 0.42 & 0.78 & 0.33 \\
Others                                    & 0.05                 & 0.06 & 0.07 & 0.07 & 0.10 & 0.07    \Bstrut   \\   \hline
\end{tabular}
\end{threeparttable}}
\end{table}

\begin{table}[H]
\caption{\textbf{Average Equation Size Distribution for Blocks in the Residual Covariance Matrix (5-Factor VHAR).}}
\label{table_average_size_block_5f}
\centering
\begin{minipage}{\linewidth}
\begin{footnotesize}
The distribution represents the percentiles and means values calculated over 473 estimated models. The 430 stocks are separated into ten groups, according to SIC. Averages are calculated among stocks in the same group.
\end{footnotesize}
\end{minipage}
\resizebox{\columnwidth}{!}{
\begin{threeparttable}
\begin{tabular}{lcccccc}  \hline \Tstrut
                  Distribution                        & \multicolumn{1}{l}{Min} & \multicolumn{1}{l}{0,25} & \multicolumn{1}{l}{0,5} & \multicolumn{1}{l}{0,75} & \multicolumn{1}{l}{Max} & \multicolumn{1}{l}{Mean} \Bstrut \\  \hline
                                          & \multicolumn{1}{l}{}    & \multicolumn{1}{l}{}     & \multicolumn{1}{l}{}    & \multicolumn{1}{l}{}     & \multicolumn{1}{l}{}    & \multicolumn{1}{l}{}  \\
                                          & \multicolumn{6}{c}{Variance Equations}                                                                                                                       \\  \cline{2-7}
\textit{Group}                            & \multicolumn{1}{l}{}    & \multicolumn{1}{l}{}     & \multicolumn{1}{l}{}    & \multicolumn{1}{l}{}     & \multicolumn{1}{l}{}    & \multicolumn{1}{l}{}     \\
Consumer Non-Durables                     & 11.32                                    & 11.97                    & 12.19                   & 12.45                    & 13.61                   & 12.27                    \\
Consumer Durables                         & 6.12                                     & 6.50                     & 6.62                    & 7.00                     & 7.62                    & 6.77                     \\
Manufacturing                             & 15.91                                    & 16.46                    & 16.72                   & 16.97                    & 17.57                   & 16.72                    \\
Oil, Gas, and Coal Extraction             & 10.28                                    & 10.75                    & 11.12                   & 11.38                    & 12.06                   & 11.08                    \\
Business Equipment                        & 15.00                                    & 15.34                    & 15.52                   & 15.72                    & 16.39                   & 15.55                    \\
Telecommunications                        & 6.50                                     & 6.90                     & 7.30                    & 7.70                     & 7.90                    & 7.30                     \\
Wholesale and Retail                      & 12.56                                    & 13.22                    & 13.40                   & 13.62                    & 14.24                   & 13.42                    \\
Health Care, Medical Equipment, and Drugs & 11.69                                    & 12.15                    & 12.54                   & 12.96                    & 13.50                   & 12.55                    \\
Utilities                                 & 13.47                                    & 14.22                    & 14.36                   & 14.56                    & 14.97                   & 14.37                    \\
Others                                    & 15.35                                    & 15.94                    & 16.60                   & 17.22                    & 18.12                   & 16.61                    \\
                                          & \multicolumn{1}{l}{}    & \multicolumn{1}{l}{}     & \multicolumn{1}{l}{}    & \multicolumn{1}{l}{}     & \multicolumn{1}{l}{}    & \multicolumn{1}{l}{}     \\
                                          & \multicolumn{6}{c}{Covariance Equations}                                                                                                                     \\   \cline{2-7}
\textit{Group}                            & \multicolumn{1}{l}{}    & \multicolumn{1}{l}{}     & \multicolumn{1}{l}{}    & \multicolumn{1}{l}{}     & \multicolumn{1}{l}{}    & \multicolumn{1}{l}{}     \\
Consumer Non-Durables                     & 0.59                                     & 0.83                     & 1.33                    & 1.43                     & 1.52                    & 1.17                     \\
Consumer Durables                         & 0.21                                     & 0.39                     & 0.46                    & 0.57                     & 0.79                    & 0.48                     \\
Manufacturing                             & 0.46                                     & 0.53                     & 0.61                    & 0.63                     & 0.69                    & 0.59                     \\
Oil, Gas, and Coal Extraction             & 1.38                                     & 1.67                     & 1.73                    & 1.95                     & 2.33                    & 1.80                     \\
Business Equipment                        & 0.37                                     & 0.42                     & 0.57                    & 0.68                     & 0.74                    & 0.56                     \\
Telecommunications                        & 1.13                                     & 1.44                     & 1.56                    & 1.69                     & 1.91                    & 1.55                     \\
Wholesale and Retail                      & 0.59                                     & 0.70                     & 0.97                    & 1.06                     & 1.13                    & 0.90                     \\
Health Care, Medical Equipment, and Drugs & 0.52                                     & 0.75                     & 1.20                    & 1.54                     & 1.78                    & 1.16                     \\
Utilities                                 & 0.64                                     & 0.74                     & 0.95                    & 1.15                     & 1.39                    & 0.95                     \\
Others                                    & 0.60                                     & 0.67                     & 0.70                    & 0.74                     & 0.77                    & 0.70     \Bstrut   \\   \hline
\end{tabular}
\end{threeparttable}}
\end{table}

\begin{table}[H]
\caption{\textbf{Average Parameter Change Distribution for Blocks in the Residual Covariance Matrix (5-Factor VHAR).}}
\label{table_average_change_block_5f}
\centering
\begin{minipage}{\linewidth}
\begin{footnotesize}
The distribution represents the percentiles and means values calculated over 473 estimated models. The 430 stocks are separated into ten groups, according to SIC. Averages are calculated among stocks in the same group.
\end{footnotesize}
\end{minipage}
\resizebox{\columnwidth}{!}{
\begin{threeparttable}
\bigskip
\begin{tabular}{lcccccc}  \hline \Tstrut
                  Distribution                        & \multicolumn{1}{l}{Min} & \multicolumn{1}{l}{0,25} & \multicolumn{1}{l}{0,5} & \multicolumn{1}{l}{0,75} & \multicolumn{1}{l}{Max} & \multicolumn{1}{l}{Mean} \Bstrut \\  \hline
                                          & \multicolumn{1}{l}{}    & \multicolumn{1}{l}{}     & \multicolumn{1}{l}{}    & \multicolumn{1}{l}{}     & \multicolumn{1}{l}{}    & \multicolumn{1}{l}{}  \\
                                          & \multicolumn{6}{c}{Variance Equations}                                                                                                                       \\  \cline{2-7}
\textit{Group}                            & \multicolumn{1}{l}{}    & \multicolumn{1}{l}{}     & \multicolumn{1}{l}{}    & \multicolumn{1}{l}{}     & \multicolumn{1}{l}{}    & \multicolumn{1}{l}{}     \\
Consumer Non-Durables                     & 0.10                 & 0.83 & 1.25 & 1.56 & 3.95 & 1.27 \\
Consumer Durables                         & 0.00                 & 0.00 & 0.00 & 0.00 & 3.12 & 0.41 \\
Manufacturing                             & 0.31                 & 0.88 & 1.07 & 1.30 & 3.98 & 1.12 \\
Oil, Gas, and Coal Extraction             & 0.10                 & 0.59 & 0.88 & 1.27 & 4.49 & 0.97 \\
Business Equipment                        & 0.27                 & 0.97 & 1.16 & 1.37 & 3.55 & 1.19 \\
Telecommunications                        & 0.00                 & 0.00 & 0.00 & 1.00 & 6.00 & 0.61 \\
Wholesale and Retail                      & 0.20                 & 0.94 & 1.23 & 1.53 & 4.10 & 1.27 \\
Health Care, Medical Equipment, and Drugs & 0.00                 & 0.59 & 1.04 & 1.63 & 4.44 & 1.17 \\
Utilities                                 & 0.00                 & 0.85 & 1.16 & 1.47 & 5.71 & 1.19 \\
Others                                    & 0.42                 & 0.66 & 0.77 & 0.91 & 1.66 & 0.81 \\
                                          & \multicolumn{1}{l}{}    & \multicolumn{1}{l}{}     & \multicolumn{1}{l}{}    & \multicolumn{1}{l}{}     & \multicolumn{1}{l}{}    & \multicolumn{1}{l}{}     \\
                                          & \multicolumn{6}{c}{Covariance Equations}                                                                                                                     \\   \cline{2-7}
\textit{Group}                            & \multicolumn{1}{l}{}    & \multicolumn{1}{l}{}     & \multicolumn{1}{l}{}    & \multicolumn{1}{l}{}     & \multicolumn{1}{l}{}    & \multicolumn{1}{l}{}     \\
Consumer Non-Durables                     & 0.10                 & 0.26 & 0.33 & 0.41 & 0.68 & 0.33 \\
Consumer Durables                         & 0.00                 & 0.00 & 0.00 & 0.45 & 3.57 & 0.41 \\
Manufacturing                             & 0.05                 & 0.09 & 0.10 & 0.12 & 0.17 & 0.10 \\
Oil, Gas, and Coal Extraction             & 0.14                 & 0.42 & 0.49 & 0.58 & 0.95 & 0.50 \\
Business Equipment                        & 0.04                 & 0.08 & 0.10 & 0.12 & 0.18 & 0.10 \\
Telecommunications                        & 0.00                 & 0.22 & 0.67 & 0.89 & 3.11 & 0.68 \\
Wholesale and Retail                      & 0.08                 & 0.16 & 0.20 & 0.24 & 0.39 & 0.20 \\
Health Care, Medical Equipment, and Drugs & 0.07                 & 0.26 & 0.39 & 0.50 & 1.11 & 0.39 \\
Utilities                                 & 0.08                 & 0.20 & 0.28 & 0.36 & 0.59 & 0.28 \\
Others                                    & 0.05                 & 0.07 & 0.07 & 0.08 & 0.10 & 0.07     \Bstrut   \\   \hline
\end{tabular}
\end{threeparttable}}
\end{table}

\begin{table}[H]
\caption{\textbf{Average Equation Size Distribution for Blocks in the Residual Covariance Matrix (7-Factor HVAR).}}
\label{table_average_size_block_7f}
\centering
\begin{minipage}{\linewidth}
\begin{footnotesize}
The distribution represents the percentiles and mean value calculated over 473 estimated models. The 430 stocks are separated into 10 groups, according to SIC. Averages are calculated among stocks in the same group.
\end{footnotesize}
\end{minipage}
\resizebox{\columnwidth}{!}{
\begin{threeparttable}
\bigskip
\begin{tabular}{lcccccc}  \hline \Tstrut
                  Distribution                        & \multicolumn{1}{l}{Min} & \multicolumn{1}{l}{0,25} & \multicolumn{1}{l}{0,5} & \multicolumn{1}{l}{0,75} & \multicolumn{1}{l}{Max} & \multicolumn{1}{l}{Mean} \Bstrut \\  \hline
                                          & \multicolumn{1}{l}{}    & \multicolumn{1}{l}{}     & \multicolumn{1}{l}{}    & \multicolumn{1}{l}{}     & \multicolumn{1}{l}{}    & \multicolumn{1}{l}{}  \\
                                          & \multicolumn{6}{c}{Variance Equations}                                                                                                                       \\  \cline{2-7}
\textit{Group}                            & \multicolumn{1}{l}{}    & \multicolumn{1}{l}{}     & \multicolumn{1}{l}{}    & \multicolumn{1}{l}{}     & \multicolumn{1}{l}{}    & \multicolumn{1}{l}{}     \\
Consumer Non-Durables                     & 11.29                & 11.94 & 12.13 & 12.48 & 13.74 & 12.24 \\
Consumer Durables                         & 6.12                 & 6.38  & 6.50  & 6.88  & 7.62  & 6.65  \\
Manufacturing                             & 15.82                & 16.38 & 16.65 & 16.91 & 17.37 & 16.65 \\
Oil, Gas, and Coal Extraction             & 10.12                & 10.81 & 11.03 & 11.34 & 11.97 & 11.09 \\
Business Equipment                        & 14.72                & 15.20 & 15.38 & 15.56 & 16.05 & 15.37 \\
Telecommunications                        & 6.40                 & 6.80  & 7.30  & 7.50  & 8.00  & 7.23  \\
Wholesale and Retail                      & 12.73                & 13.31 & 13.47 & 13.69 & 14.33 & 13.49 \\
Health Care, Medical Equipment, and Drugs & 11.77                & 12.31 & 12.69 & 13.15 & 13.73 & 12.73 \\
Utilities                                 & 13.69                & 14.22 & 14.36 & 14.53 & 14.92 & 14.35 \\
Others                                    & 15.36                & 15.93 & 16.57 & 17.12 & 18.05 & 16.59 \\
                                          & \multicolumn{1}{l}{}    & \multicolumn{1}{l}{}     & \multicolumn{1}{l}{}    & \multicolumn{1}{l}{}     & \multicolumn{1}{l}{}    & \multicolumn{1}{l}{}     \\
                                          & \multicolumn{6}{c}{Covariance Equations}                                                                                                                     \\   \cline{2-7}
\textit{Group}                            & \multicolumn{1}{l}{}    & \multicolumn{1}{l}{}     & \multicolumn{1}{l}{}    & \multicolumn{1}{l}{}     & \multicolumn{1}{l}{}    & \multicolumn{1}{l}{}     \\
Consumer Non-Durables                     & 0.59                 & 0.82  & 1.33  & 1.41  & 1.51  & 1.15  \\
Consumer Durables                         & 0.14                 & 0.36  & 0.46  & 0.54  & 0.75  & 0.45  \\
Manufacturing                             & 0.50                 & 0.56  & 0.63  & 0.65  & 0.72  & 0.61  \\
Oil, Gas, and Coal Extraction             & 1.40                 & 1.64  & 1.77  & 2.05  & 2.42  & 1.84  \\
Business Equipment                        & 0.54                 & 0.62  & 0.77  & 0.90  & 0.95  & 0.76  \\
Telecommunications                        & 1.58                 & 1.82  & 1.91  & 2.02  & 2.56  & 1.93  \\
Wholesale and Retail                      & 0.62                 & 0.73  & 1.03  & 1.11  & 1.16  & 0.93  \\
Health Care, Medical Equipment, and Drugs & 0.54                 & 0.77  & 1.02  & 1.41  & 1.65  & 1.06  \\
Utilities                                 & 0.78                 & 0.88  & 1.06  & 1.12  & 1.25  & 1.00  \\
Others                                    & 0.63                 & 0.71  & 0.75  & 0.78  & 0.81  & 0.74   \Bstrut   \\   \hline
\end{tabular}
\end{threeparttable}}
\end{table}

\begin{table}[H]
\caption{\textbf{Average Parameter Change Distribution for Blocks in the Residual Covariance Matrix (7-Factor HVAR).}}
\label{table_average_change_block_7f}
\centering
\begin{minipage}{\linewidth}
\begin{footnotesize}
The distribution represents the percentiles and means values calculated over 473 estimated models. The 430 stocks are separated into ten groups, according to SIC. Averages are calculated among stocks in the same group.
\end{footnotesize}
\end{minipage}
\resizebox{\columnwidth}{!}{
\begin{threeparttable}
\bigskip
\begin{tabular}{lcccccc}  \hline \Tstrut
                  Distribution                        & \multicolumn{1}{l}{Min} & \multicolumn{1}{l}{0.25} & \multicolumn{1}{l}{0.5} & \multicolumn{1}{l}{0.75} & \multicolumn{1}{l}{Max} & \multicolumn{1}{l}{Mean} \Bstrut \\  \hline
                                          & \multicolumn{1}{l}{}    & \multicolumn{1}{l}{}     & \multicolumn{1}{l}{}    & \multicolumn{1}{l}{}     & \multicolumn{1}{l}{}    & \multicolumn{1}{l}{}  \\
                                          & \multicolumn{6}{c}{Variance Equations}                                                                                                                       \\  \cline{2-7}
\textit{Group}                            & \multicolumn{1}{l}{}    & \multicolumn{1}{l}{}     & \multicolumn{1}{l}{}    & \multicolumn{1}{l}{}     & \multicolumn{1}{l}{}    & \multicolumn{1}{l}{}     \\
Consumer Non-Durables                     & 0.00                 & 0.94 & 1.25 & 1.66 & 3.64 & 1.32 \\
Consumer Durables                         & 0.00                 & 0.00 & 0.00 & 0.00 & 3.12 & 0.40 \\
Manufacturing                             & 0.38                 & 0.88 & 1.04 & 1.25 & 3.74 & 1.10 \\
Oil, Gas, and Coal Extraction             & 0.00                 & 0.59 & 0.88 & 1.17 & 3.71 & 0.92 \\
Business Equipment                        & 0.46                 & 0.91 & 1.13 & 1.35 & 3.14 & 1.16 \\
Telecommunications                        & 0.00                 & 0.00 & 0.00 & 1.00 & 8.00 & 0.67 \\
Wholesale and Retail                      & 0.25                 & 0.89 & 1.19 & 1.53 & 3.90 & 1.24 \\
Health Care, Medical Equipment, and Drugs & 0.00                 & 0.59 & 1.04 & 1.63 & 4.44 & 1.19 \\
Utilities                                 & 0.23                 & 0.85 & 1.16 & 1.47 & 6.10 & 1.22 \\
Others                                    & 0.36                 & 0.65 & 0.77 & 0.92 & 1.73 & 0.80 \\
                                          & \multicolumn{1}{l}{}    & \multicolumn{1}{l}{}     & \multicolumn{1}{l}{}    & \multicolumn{1}{l}{}     & \multicolumn{1}{l}{}    & \multicolumn{1}{l}{}     \\
                                          & \multicolumn{6}{c}{Covariance Equations}                                                                                                                     \\   \cline{2-7}
\textit{Group}                            & \multicolumn{1}{l}{}    & \multicolumn{1}{l}{}     & \multicolumn{1}{l}{}    & \multicolumn{1}{l}{}     & \multicolumn{1}{l}{}    & \multicolumn{1}{l}{}     \\
Consumer Non-Durables                     & 0.05                 & 0.25 & 0.33 & 0.40 & 0.65 & 0.33 \\
Consumer Durables                         & 0.00                 & 0.00 & 0.00 & 0.89 & 2.23 & 0.43 \\
Manufacturing                             & 0.06                 & 0.09 & 0.10 & 0.12 & 0.18 & 0.11 \\
Oil, Gas, and Coal Extraction             & 0.17                 & 0.43 & 0.51 & 0.61 & 0.92 & 0.52 \\
Business Equipment                        & 0.06                 & 0.11 & 0.12 & 0.14 & 0.25 & 0.13 \\
Telecommunications                        & 0.00                 & 0.44 & 0.89 & 1.33 & 4.00 & 0.93 \\
Wholesale and Retail                      & 0.08                 & 0.17 & 0.20 & 0.24 & 0.40 & 0.20 \\
Health Care, Medical Equipment, and Drugs & 0.06                 & 0.25 & 0.34 & 0.44 & 0.75 & 0.35 \\
Utilities                                 & 0.10                 & 0.22 & 0.29 & 0.36 & 0.63 & 0.30 \\
Others                                    & 0.05                 & 0.07 & 0.08 & 0.08 & 0.10 & 0.08   \Bstrut   \\   \hline
\end{tabular}
\end{threeparttable}}
\end{table}

\pagebreak

\section{Minimum Variance Portfolio: Partial Rebalancing}\label{partialrebalancing}

This section shows the return statistics for cases in which we do not fully rebalance the portfolio daily. We assume that the investor updates only 1/22 of the portfolio each day based on the predicted covariance matrix for the following day, implying that every daily portfolio update has an effective one-month holding period. This change accounts for the possible effect of transaction costs, as only a fraction of the portfolio changes daily. Indirectly, we also test whether investors with a one-month horizon could benefit from using covariance estimates from the proposed model, which are designed to forecast one-day ahead covariance matrices.

Tables \ref{table_minimum_variance_agg}, \ref{table_restricted_portfolio_agg}, and \ref{table_restricted_portfolio_long_only_agg} present results for the unrestricted, restricted, and long-only cases, respectively. Our proposed model performs better than the alternatives, even when we focus on longer holding periods.

\begin{landscape}
\begin{table}[H]
\caption{\textbf{Statistics for Daily Average Portfolios - Global Minimum Variance.}}
\label{table_minimum_variance_agg}
\begin{minipage}{\linewidth}
\begin{footnotesize}
The table reports the results for portfolios constructed according to the optimization problem as in (\ref{eq_minimum_variance}) with partial rebalancing. Different models provide the one-step-ahead forecasts for the realized covariance matrix: RW is the random walk model, and Blocks 1F, 3F, 5F, and 7F are random walks applied to the residual covariance matrix after the four different factor decompositions.
\end{footnotesize}
\end{minipage}
\begin{threeparttable}
\bigskip
\resizebox{\linewidth}{!}{
\begin{tabular}{ccccccccccc}
\hline

                                      & RW & Block 1F & Block 3F & Block 5F & Block 7F & EWMA (Returns)& BEKK-NL & DCC-NL & AFM1-DCC-NL & IDR-DCC-NL \\
                                      \hline
Standard Deviation (\%)               & 11.08             & 9.49              & 9.57              & 9.40              & 9.43              & 14.00             & 9.58              & 10.64   &  9.15    & 10.88 \\
Lower Partial Standard Deviation (\%) & 12.57             & 10.69             & 10.75             & 10.48             & 10.50             & 14.16             & 9.65              & 11.09   &   9.41   & 11.43 \\
Kurtosis                              & 4.86              & 5.11              & 4.76              & 4.66              & 4.94              & 2.09              & 1.57              & 4.08    & 2.71     & 4.88  \\
Skewness                              & --0.72            & --0.88            & --0.85            & --0.81            & --0.84            & --0.04            & --0.18            & --0.09  & 0.01     & -0.24 \\
Average Diversification Ratio         & 7.09              & 7.19              & 7.29              & 7.30              & 7.35              & 1.08              & 3.07              & 3.90    & 3.32     & 4.13  \\
Average Max. Weight                   & 0.03              & 0.07              & 0.07              & 0.07              & 0.07              & 0.19              & 0.06              & 0.10    & 0.12     & 0.11  \\
Average Min. Weight                   & --0.02            & --0.01            & --0.01            & --0.01            & --0.01            & --0.14            & --0.04            & --0.03  & --0.05   & -0.02 \\
Average Gross Leverage                & 2.57              & 2.14              & 2.15              & 2.15              & 2.16              & 11.89             & 5.01              & 3.74    & 4.30     & 3.04   \\
Proportion of Leverage (\%)           & 37.74             & 42.09             & 41.75             & 41.36             & 41.54             & 49.07             & 45.15             & 50.94   & 47.10    & 48.4   \\
Average Turnover (\%)                 & 0.09              & 0.04              & 0.04              & 0.04              & 0.04              & 0.08              & 0.03              & 0.04    &  0.03    & 0.04   \\
Average Excess Return (\%)            & 17.06             & 14.41             & 15.00             & 15.61             & 15.36             & 3.73              & 17.52             & 15.50   & 14.71    & 17.34  \\
Cumulative Return (\%)                & 36.44             & 30.21             & 31.65             & 33.21             & 32.56             & 5.52              & 38.03             & 32.61   & 31.04    & 37.2   \\
Sharpe Ratio                          & 1.54              & 1.52              & 1.57              & 1.66              & 1.63              & 0.27              & 1.83              & 1.46    &  1.61    & 1.59   \\
\hline
& \multicolumn{2}{c}{1 Factor}     & \multicolumn{2}{c}{3 Factors}  &    \multicolumn{2}{c}{5 Factors}    & \multicolumn{2}{c}{7 Factors}    \\
& \multicolumn{2}{c}{VHAR}         & \multicolumn{2}{c}{VHAR}    &      \multicolumn{2}{c}{VHAR}         & \multicolumn{2}{c}{VHAR}         \\
& \multicolumn{2}{c}{(Log matrix)} & \multicolumn{2}{c}{(Log matrix)}   & \multicolumn{2}{c}{(Log matrix)} & \multicolumn{2}{c}{(Log matrix)} \\
& LASSO             & adaLASSO          & LASSO             & adaLASSO          & LASSO             & adaLASSO          & LASSO             & adaLASSO \\
\hline
Standard Deviation (\%)               & 9.13              & 9.11              & 9.09              & 9.06              & 8.98              & 8.96              & 8.86              & 8.84              \\
Lower Partial Standard Deviation (\%) & 9.84              & 9.81              & 9.79              & 9.77              & 9.72              & 9.72              & 9.72              & 9.76              \\
Kurtosis                              & 2.76              & 2.83              & 2.37              & 2.40              & 2.18              & 2.21              & 2.72              & 2.74              \\
Skewness                              & --0.53             & --0.53             & --0.50             & --0.50           & --0.47             & --0.47             & --0.58             & --0.58             \\
Average Diversification Ratio         & 4.84              & 4.88              & 5.05              & 5.08              & 4.85              & 4.88              & 4.96              & 4.98              \\
Average Max. Weight                   & 0.06              & 0.07              & 0.07              & 0.07             & 0.07              & 0.08              & 0.07              & 0.08              \\
Average Min. Weight                   & --0.01             & --0.01             & --0.02             & --0.02            & --0.02             & --0.02             & --0.02             & --0.02             \\
Average Gross Leverage                & 2.56              & 2.56              & 2.70              & 2.70              & 2.74              & 2.74              & 2.84              & 2.84              \\
Proportion of Leverage (\%)           & 45.61             & 45.70             & 44.64             & 44.78            & 44.65             & 44.81             & 44.95             & 45.14             \\
Average Turnover (\%)                 & 0.02              & 0.02              & 0.02              & 0.02              & 0.02              & 0.02              & 0.02              & 0.02              \\
Average Excess Return (\%)            & 13.75             & 13.56             & 15.95             & 15.66             & 16.70             & 16.30             & 15.60             & 15.33             \\
Cumulative Return (\%)                & 28.69             & 28.23             & 34.12             & 33.40             & 36.05             & 35.05             & 33.30             & 32.63             \\
Sharpe Ratio                          & 1.51              & 1.49              & 1.75              & 1.73              & 1.86              & 1.82              & 1.76              & 1.73       \\
\hline
\end{tabular}}
\end{threeparttable}
\end{table}
\end{landscape}

\begin{landscape}
\begin{table}[H]
\caption{\textbf{Statistics for Daily Average Portfolios - Restricted Minimum Variance.}}
\label{table_restricted_portfolio_agg}
\begin{minipage}{\linewidth}
\begin{footnotesize}
The table reports the results for portfolios constructed according to the optimization problem as in (\ref{eq_minimum_variance_restricted}) with partial rebalancing. Different models provide the one-step-ahead forecasts for the realized covariance matrix: RW is the random walk model, and Blocks 1F, 3F, 5F, and 7F are random walks applied to the residual covariance matrix after the four different factor decompositions.\end{footnotesize}
\end{minipage}
\begin{threeparttable}
\bigskip
\resizebox{\linewidth}{!}{
\begin{tabular}{ccccccccccc}
\hline

                                      & RW & Block 1F & Block 3F & Block 5F & Block 7F & EWMA (Returns)& BEKK-NL & DCC-NL & AFM1-DCC-NL & IDR-DCC-NL \\
                                      \hline
Standard Deviation (\%)               & 13.10             & 12.79             & 12.62             & 12.56             & 12.63             & 14.80             & 15.01             & 14.55    & 15.92  & 13.27      \\
Lower Partial Standard Deviation (\%) & 13.85             & 13.34             & 13.19             & 13.03             & 13.16             & 15.53             & 15.58             & 15.40    & 17.60  & 13.81      \\
Kurtosis                              & 3.33              & 3.10              & 3.27              & 3.29              & 3.40              & 3.35              & 3.26              & 3.14     & 2.17   & 3.39      \\
Skewness                              & --0.37            & --0.31            & --0.35            & --0.34            & --0.36            & --0.35            & --0.28            & --0.29   & --0.50 & --0.34      \\
Average Diversification Ratio         & 4.61              & 4.53              & 4.78              & 4.78              & 4.78              & 2.38              & 2.44              & 2.35     & 1.66   & 2.89      \\
Average Max. Weight                   & 0.12              & 0.11              & 0.11              & 0.11              & 0.11              & 0.18              & 0.19              & 0.19     & 0.19   & 0.18      \\
Average Min. Weight                   & --0.05            & --0.05            & --0.05            & --0.05            & --0.05            & --0.15            & --0.14            & --0.10   & --0.17 & --0.04      \\
Average Gross Leverage                & 1.43              & 1.48              & 1.48              & 1.48              & 1.48              & 1.57              & 1.56              & 1.55     & 1.60   & 1.51      \\
Proportion of Leverage (\%)           & 7.63              & 14.89             & 14.51             & 13.71             & 13.26             & 1.16              & 1.64              & 2.75     & 0.83   & 10.34      \\
Average Turnover (\%)                 & 0.02              & 0.02              & 0.02              & 0.02              & 0.02              & 0.01              & 0.01              & 0.02     & 0.01   & 0.02      \\
Average Excess Return (\%)            & 16.75             & 18.48             & 18.64             & 18.36             & 17.74             & 15.38             & 14.05             & 13.75    & 9.75   & 18.66      \\
Cumulative Return (\%)                & 35.03             & 39.58             & 40.07             & 39.34             & 37.72             & 31.01             & 27.71             & 27.17    & 17.49  & 39.88      \\
Sharpe Ratio                          & 1.28              & 1.44              & 1.48              & 1.46              & 1.41              & 1.04              & 0.94              & 0.95     & 0.61   & 1.41      \\
\hline
& \multicolumn{2}{c}{1 Factor}     & \multicolumn{2}{c}{3 Factors}  &   \multicolumn{2}{c}{5 Factors}    & \multicolumn{2}{c}{7 Factors}    \\
& \multicolumn{2}{c}{VHAR}         & \multicolumn{2}{c}{VHAR}    &      \multicolumn{2}{c}{VHAR}         & \multicolumn{2}{c}{VHAR}         \\
& \multicolumn{2}{c}{(Log matrix)} & \multicolumn{2}{c}{(Log matrix)} &  \multicolumn{2}{c}{(Log matrix)} & \multicolumn{2}{c}{(Log matrix)} \\
& LASSO             & adaLASSO          & LASSO             & adaLASSO          & LASSO             & adaLASSO          & LASSO             & adaLASSO \\
\hline
Standard Deviation (\%)               & 12.68             & 12.74             & 12.21             & 12.35             & 12.07             & 12.17             & 12.19             & 12.24             \\
Lower Partial Standard Deviation (\%) & 12.94             & 12.95             & 12.40             & 12.57             & 12.41             & 12.43             & 12.67             & 12.72             \\
Kurtosis                              & 2.59              & 2.73              & 2.96              & 2.99              & 3.05              & 3.04              & 3.19              & 3.03              \\
Skewness                              & --0.23             & --0.24             & --0.24             & --0.26             & --0.27             & --0.25             & --0.30             & --0.26             \\
Average Diversification Ratio         & 3.62              & 3.61              & 3.87              & 3.86              & 3.91              & 3.91              & 3.87              & 3.85              \\
Average Max. Weight                   & 0.13              & 0.13              & 0.13              & 0.13              & 0.13              & 0.13              & 0.13              & 0.13              \\
Average Min. Weight                   & --0.07             & --0.07             & --0.06             & --0.06             & --0.06             & --0.06             & --0.06             & --0.06             \\
Average Gross Leverage                & 1.58              & 1.58              & 1.59              & 1.59              & 1.59              & 1.59              & 1.59              & 1.59              \\
Proportion of Leverage (\%)           & 8.52              & 8.71              & 7.72              & 7.97              & 7.68              & 7.85              & 7.12              & 7.34              \\
Average Turnover (\%)                 & 0.02              & 0.02              & 0.02              & 0.02              & 0.02              & 0.02              & 0.02              & 0.02              \\
Average Excess Return (\%)            & 18.12             & 17.87             & 17.94             & 18.10             & 17.65             & 17.97             & 17.20             & 17.42             \\
Cumulative Return (\%)                & 38.69             & 38.02             & 38.37             & 38.74             & 37.65             & 38.45             & 36.47             & 37.00             \\
Sharpe Ratio                          & 1.43              & 1.40              & 1.47              & 1.47              & 1.46              & 1.48              & 1.41              & 1.42     \\
\hline
\end{tabular}}
\end{threeparttable}
\end{table}
\end{landscape}

\begin{landscape}
\begin{table}[H]
\caption{\textbf{Statistics for Daily Average Portfolios - Restricted Minimum Variance (Long Only).}}
\label{table_restricted_portfolio_long_only_agg}
\begin{minipage}{\linewidth}
\begin{footnotesize}
The table reports the results for portfolios constructed according to the optimization problem as in (\ref{eq_minimum_variance_long_only}) with partial rebalancing. Different models provide the one-step-ahead forecasts for the realized covariance matrix: RW is the random walk model, and Blocks 1F, 3F, 5F, and 7F are random walks applied to the residual covariance matrix after the four different factor decompositions.
\end{footnotesize}
\end{minipage}
\begin{threeparttable}
\bigskip
\resizebox{\linewidth}{!}{
\begin{tabular}{ccccccccccc}
\hline

                                      & RW & Block 1F & Block 3F & Block 5F & Block 7F & EWMA (Returns)& BEKK-NL & DCC - NL & AFM1-DCC-NL\\
                                      \hline
Standard Deviation (\%)               & 16.66             & 16.73             & 16.60             & 16.49             & 16.51             & 17.53             & 17.73             & 17.58     & 20.66 & 17.4      \\
Lower Partial Standard Deviation (\%) & 17.85             & 17.69             & 17.45             & 17.40             & 17.49             & 18.56             & 18.78             & 18.62     & 21.41 & 18.44      \\
Kurtosis                              & 2.70              & 2.76              & 2.79              & 2.78              & 2.83              & 2.78              & 2.53              & 2.63      & 2.22  & 3  \\
Skewness                              & --0.33             & --0.32             & --0.32             & --0.32             & --0.33             & --0.32             & --0.31             & --0.29     &  --0.35& --0.29      \\
Average Diversification Ratio         & 3.37              & 3.27              & 3.30              & 3.32              & 3.32              & 2.72              & 2.69              & 2.61      & 2.18  & 2.81      \\
Average Max. Weight                   & 0.15              & 0.17              & 0.17              & 0.17              & 0.17              & 0.17              & 0.17              & 0.17      & 0.13  & 0.18     \\
Average Min. Weight                   & 0.00              & 0.00              & 0.00              & 0.00              & 0.00              & 0.00              & 0.00              & 0.00      & 0.00  & 0    \\
Average Gross Leverage                & 1.00              & 1.00              & 1.00              & 1.00              & 1.00              & 1.00              & 1.00              & 1.00      & 1.00  & 1     \\
Proportion of Leverage (\%)           & 0.00              & 0.00              & 0.00              & 0.00              & 0.00              & 0.00              & 0.00              & 0.00      & 0.00  & 0       \\
Average Turnover (\%)                 & 0.01              & 0.01              & 0.01              & 0.01              & 0.01              & 0.01              & 0.01              & 0.01      & 0.01  & 0.01     \\
Average Excess Return (\%)            & 17.41             & 16.96             & 17.35             & 17.36             & 17.38             & 17.74             & 16.90             & 16.28     & 14.58 & 14.92         \\
Cumulative Return (\%)                & 35.36             & 34.18             & 35.23             & 35.30             & 35.35             & 35.81             & 33.61             & 32.12     & 26.56 & 28.87      \\
Sharpe Ratio                          & 1.05              & 1.01              & 1.05              & 1.05              & 1.05              & 1.01              & 0.95              & 0.93      & 0.71  & 0.86   \\
\hline
& \multicolumn{2}{c}{1 Factor}     & \multicolumn{2}{c}{3 Factors} &    \multicolumn{2}{c}{5 Factors}    & \multicolumn{2}{c}{7 Factors}    \\
& \multicolumn{2}{c}{VHAR}         & \multicolumn{2}{c}{VHAR}    &      \multicolumn{2}{c}{VHAR}         & \multicolumn{2}{c}{VHAR}         \\
& \multicolumn{2}{c}{(Log matrix)} & \multicolumn{2}{c}{(Log matrix)} &    \multicolumn{2}{c}{(Log matrix)} & \multicolumn{2}{c}{(Log matrix)} \\
& LASSO             & adaLASSO          & LASSO             & adaLASSO     &     LASSO             & adaLASSO          & LASSO             & adaLASSO \\
\hline
Standard Deviation (\%)               & 16.66             & 16.73             & 16.60             & 16.49             & 16.51             & 17.53             & 17.73             & 17.58             \\
Lower Partial Standard Deviation (\%) & 17.85             & 17.69             & 17.45             & 17.40             & 17.49             & 18.56             & 18.78             & 18.62             \\
Kurtosis                              & 2.70              & 2.76              & 2.79              & 2.78              & 2.83              & 2.78              & 2.53              & 2.63              \\
Skewness                              & --0.33             & --0.32             & --0.32             & --0.32             & --0.33             & --0.32             & --0.31             & --0.29             \\
Average Diversification Ratio         & 3.37              & 3.27              & 3.30              & 3.32              & 3.32              & 2.72              & 2.69              & 2.61              \\
Average Max. Weight                   & 0.15              & 0.17              & 0.17              & 0.17              & 0.17              & 0.17              & 0.17              & 0.17              \\
Average Min. Weight                   & 0.00              & 0.00              & 0.00              & 0.00              & 0.00              & 0.00              & 0.00              & 0.00              \\
Average Gross Leverage                & 1.00              & 1.00              & 1.00              & 1.00              & 1.00              & 1.00              & 1.00              & 1.00              \\
Proportion of Leverage (\%)           & 0.00              & 0.00              & 0.00              & 0.00              & 0.00              & 0.00              & 0.00              & 0.00              \\
Average Turnover (\%)                 & 0.01              & 0.01              & 0.01              & 0.01              & 0.01              & 0.01              & 0.01              & 0.01              \\
Average Excess Return (\%)            & 17.41             & 16.96             & 17.35             & 17.36             & 17.38             & 17.74             & 16.90             & 16.28             \\
Cumulative Return (\%)                & 35.36             & 34.18             & 35.23             & 35.30             & 35.35             & 35.81             & 33.61             & 32.12             \\
Sharpe Ratio                          & 1.05              & 1.01              & 1.05              & 1.05              & 1.05              & 1.01              & 0.95              & 0.93             \\
\hline
\end{tabular}}
\end{threeparttable}
\end{table}
\end{landscape}

\end{document}